\documentclass[10pt]{article} 
\usepackage{latexsym,amsmath,amssymb}
\usepackage{lscape,enumitem}
\usepackage{array}
\usepackage{epstopdf,caption,subcaption,graphicx,xcolor,algorithm,algorithmicx}
\usepackage[noend]{algpseudocode}
\usepackage{tikz,pgfgantt,pgfplots,needspace}
\usepackage{hyperref}
\usepackage[capitalise,noabbrev]{cleveref}
\usetikzlibrary{arrows.meta,positioning,fit,backgrounds,calc}
\usetikzlibrary{shapes,arrows,fit,calc,positioning}
\usetikzlibrary{decorations.pathreplacing}
\normalsize
\newtheorem{theorem}{Theorem}
\newtheorem{lemma}{Lemma}
\newtheorem{corollary}{Corollary}

\newtheorem{operation}{Operation}
\newtheorem{definition}{Definition}

\newtheorem{problem}{Problem}
\newenvironment{proof}{{\sc Proof. }}{\hfill$\Box$\vspace{0.1in}}
\newcommand{\OPT}{{\rm OPT}}
\newcommand{\MST}{{\rm MST}}

\newcommand{\poly}{{\rm poly}}

\def\mcC{\mathcal{C}}
\def\mbQ{\mathbb{Q}}
\def\mcK{\mathcal{K}}

\def\mcT{\mathcal{T}}
\title{Approximation algorithms for balanced spanning forest problem}
\author{
Zixin~Qiao\thanks{School of Mathematics, Hangzhou Dianzi University. Hangzhou, China.
	Emails: \texttt{\{241070003, chenyong, wx198491\} @hdu.edu.cn}}
\and
Yong~Chen$^*$\thanks{Corresponding authors.}
	\thanks{ORCID: \texttt{0000-0001-8982-7757}}
\and
Guohui~Lin\thanks{Department of Computing Science, University of Alberta.  Edmonton, Canada.
	Email: \texttt{guohui@ualberta.ca}; ORCID: \texttt{0000-0003-4283-3396}} $^{\,\dagger}$
\and
Xing~Wang$^*$
}
\date{}
\begin{document}
\maketitle

\begin{abstract}
Given an edge nonnegatively weighted connected graph $G = (V, E)$ and an integer $k \ge 2$,
the Balanced Spanning $k$-Forest problem ({\sc BS$k$F}) asks for a spanning forest consisting of at most $k$ trees that minimizes the weight of its heaviest tree.
For general graphs, a minimum spanning tree is a tight $k$-approximation for this min-max problem.
In this paper, we improve this guarantee in several regimes.
First and foremost, for every fixed $0 < \epsilon < 0.25$ and $k \ge 6/\epsilon$,
we give an $(|V|^{O(k /\epsilon)} \cdot \poly(|V|, |E|))$-time $(0.75 + \epsilon) k$-approximation algorithm.
Secondly when $k = 2$, we design a polynomial-time $1.75$-approximation algorithm.
When the input graph is complete and the edge weights satisfy triangle inequality, we design a polynomial-time $3$-approximation algorithm for any $k \ge 2$.
Our algorithm design and analyses exploit the interaction between a minimum spanning tree and an optimal spanning $k$-forest,
together with spanning-tree packing, degree-sensitive tree splitting, and a matching-based decomposition for metric complete graphs.
On the negative side, we show that {\sc BS$k$F} on metric complete graphs cannot be approximated within $1688/1687 \approx 1.00059$ unless P $=$ NP,
and this inapproximability carries over to a sparse graph class.

\end{abstract}

\section*{Acknowledgments}
The research is supported by the NNSF of China (Grants 12471301, 12401422) and
the NSERC Canada.

\newpage
\section{Introduction}
\label{sec:introduction}
Balanced graph partitioning seeks to divide a connected graph into a prescribed number of nonempty connected parts while keeping the parts as balanced as possible.
Such problems arise in applications including image processing, clustering, and multi-robot task allocation and have been studied for several decades~\cite{LPS93,MSN97,ZWD19}.

In the classical balanced connected partition setting, weights are assigned to vertices and the cost of a part is determined by the vertices it contains.
When costs are assigned to edges instead, a natural analogue is to represent each part by a spanning tree and to balance the edge cost required to interconnect the vertices in a part.
This leads to the balanced spanning-forest setting considered in this paper.

Given a connected graph $G = (V, E)$, a nonnegative edge-weight function $w: E \to \mbQ_{\ge 0}$, and an integer $k$,
the min-max balanced spanning forest problem ({\sc BS$k$F}) asks for a spanning $k$-forest such that the weight of the heaviest tree is minimized.
A feasible solution to the problem induces a partition of $V$ into $k$ vertex-disjoint connected parts,
and the objective measures the cost of the spanning tree used inside each part.
Clearly, for $k = 1$, the problem is the minimum spanning tree (MST) problem.
For $k \ge 2$, the balancing requirement creates a substantially different optimization problem,
which was first introduced by Madkour et al.~\cite{MNW19} motivated by computational topology, and models several other applications including decentralized network design~\cite{BBD19}.

The {\sc BS$k$F} problem is close in spirit to the vertex-weighted balanced connected partition problem,
but the objective is determined by the edges selected into trees rather than by the vertex weights.
It is also related to the min-max tree cover, where the cost is edge based but with an important distinction that the trees in a cover can overlap;
while our spanning forest must consist of pairwise vertex-disjoint trees.

The existing literature shows that {\sc BS$k$F} remains difficult even under strong restrictions on the instance graphs.
For $k = 2$, while the problem on trees can be solved exactly in $O(kn^3)$ time~\cite{VAP18}, where $n$ refers to the size of the tree,
Madkour et al.~\cite{MNW19} proved the NP-hardness for bipartite graphs by a reduction from the bipartition problem.
Davari et al.~\cite{DMY25} further established hardness for complete, split, and unweighted bipartite graphs, and on the positive side,
gave a tight $k$-approximation for {\sc BS$k$F} based on a minimum spanning tree of the input graph.
This leaves open whether the guarantee can be improved for small or any fixed $k$, and whether additional graph structure leads to better approximation.

One important restriction on the edge weights is triangle inequality, which maps many real-life applications.
It appears that such a restriction requires separate treatment:
Andersson et al.~\cite{AGL03} studied the {\sc BS$k$F} problem on vertex sets in the Euclidean plane,
and Guttmann-Beck and Hassin~\cite{GH97} considered an equal-cardinality metric variant.

In this paper, we study the problem in the most general settings, i.e., {\sc BS$k$F} on general graphs and on metric complete graphs,
and we study from the approximation algorithm perspective.

\subsection{Related work}
\label{subsec:related-work}
With the above overview, we roughly group the related work into four kinds.

\paragraph{Balanced connected partitions.}
The closest classical background is the vertex-weighted balanced connected partition problem.
Early complexity results were established by Dyer and Frieze~\cite{DF85},
and approximation and inapproximability results were developed by Chleb\'ikov\'a~\cite{Chl96} and Chataigner et al.~\cite{CSW07}.
For restricted graph classes, Wu~\cite{Wu12} obtained fully polynomial-time approximation schemes for a max-min variant on interval graphs when $k = 2$,
and on $k$-connected interval graphs for every fixed $k > 2$.
More recent approximation algorithms include the tripartition algorithms of Chen et al.~\cite{CCC22},
the fixed-$k$ results of Chen et al.~\cite{CCL21},
and the balanced crown decomposition of Casel et al.~\cite{CFI21} which yields constant-factor guarantees even when $k$ is part of the input.
Moura et al.~\cite{MOW23} studied approximation, parameterized algorithms, and inapproximability for balanced connected partitions,
while Miyazawa et al.~\cite{MMO21} investigated exact formulations and computational methods.
These results measure parts by vertex counts or vertex weights and therefore do not directly relate to the approximability of the edge-weighted spanning-forest objective studied here.

\paragraph{Balanced spanning forests.}
This is our line of work, and its literature is much more recent.
As mentioned earlier that Madkour et al.~\cite{MNW19} formulated the problem and studied approximation algorithms motivated by computational topology.
Vaishali et al.~\cite{VAP18} gave a polynomial-time exact algorithm on trees.
Davari et al.~\cite{DMY25} subsequently established NP-hardness on several graph classes and
developed exact approaches together with the minimum-spanning-tree based $k$-approximation algorithm,
the latter of which serves as the main baseline for this paper.
Basically, a minimum spanning tree of the input graph is computed and then its $k-1$ heaviest edges are deleted to create $k$ trees;
the approximation ratio of $k$ is shown tight for the algorithm.
Our algorithms build on the same minimum-spanning-tree but use additional information about how a minimum spanning tree crosses the vertex sets of an optimal $k$-forest.

\paragraph{Geometric and metric MST partitioning.}
Andersson et al.~\cite{AGL03} studied the problem of partitioning a set of points in the Euclidean plane into $k$ subsets
so as to minimize the maximum weight of their minimum spanning trees.
This has the same spanning-forest objective as our problem but under the special Euclidean distances.
Their geometric setting uses real random access machine arithmetic, and thus even rational coordinates could give irrational distances,
whereas our edge weights are rational and binary encoded.
Andersson et al. proved the NP-hardness and obtained a $(4/3 + \epsilon)$-approximation for $k = 2$ and a $(2 + \epsilon)$-approximation for any fixed $k \ge 3$.
Guttmann-Beck and Hassin~\cite{GH97} considered a metric complete graph with the additional requirement that all $k$ parts have equal cardinality.
They obtained approximation algorithms for this constrained variant and proved that the problem is already NP-hard for $k = 2$.
The equal-cardinality constraint makes their optimization problem different from {\sc BS$k$F},
and thus their approximation guarantees do not directly transfer to our setting.

\paragraph{Min-max tree covers.}
Min-max tree cover is another closely related edge-weighted optimization problem.
Even et al.~\cite{EGK04} obtained a $4$-approximation for the problem,
and Khani and Salavatipour~\cite{KS14} improved the factor to $3$ using a decomposition based on light and heavy components together with a matching construction.
Xu and Wen~\cite{XW10} proved an inapproximability lower bound of $3/2$ for the metric version unless P $=$ NP.
Note that an arbitrary tree cover needs not be a feasible solution for {\sc BS$k$F},
since covering trees may overlap whereas a spanning forest must induce a vertex partition.
In particular, adapting the light-heavy component and matching framework require a separate disjointness argument.

Arkin et al.~\cite{AHL06} gave a different $4$-approximation for metric tree cover based on doubling component MSTs, shortcutting, and cutting the resulting paths.
Their construction yields vertex-disjoint paths, and its analysis also applies to an optimal spanning forest.
It therefore implies a polynomial-time $4$-approximation for {\sc BS$k$F} on metric complete graphs, to be explained in Section~\ref{sec:metric-complete}.
Combining this algorithm with the MST-based one gives the previous best guarantee of $\min\{k, 4\}$ for {\sc BS$k$F} on metric complete graphs.

The literature above suggests two complementary sources of structure.
Balanced connected partition results show how connectivity constraints interact with balancing objectives,
while tree-cover algorithms demonstrate how edge-weighted metric structure can support matching and decomposition arguments.
The present work combines these perspectives with minimum-spanning-tree packing and cut structure in order to obtain guarantees specific to our target vertex-disjoint spanning forests.

\subsection{Our contributions}
\label{subsec:contributions}
We study the {\sc BS$k$F} problem from the perspectives of approximation algorithms and inapproximability, on general graphs and one metric complete graphs.

For {\sc BS$k$F} on general graphs,
we first develop an approximation framework in which the algorithm recursively recovers optimal sides separated by sufficiently small cuts of a minimum spanning tree.
If no such cut exists, the contracted graph is highly edge-connected, and so spanning-tree packing controls the total minimum-spanning-tree weight;
a balanced edge or a sink-based auxiliary construction then produces the required forest.
For every fixed $0 < \epsilon < 0.25$ and $k \ge 6/\epsilon$, this gives a $(0.75 + \epsilon) k$-approximation with running time $n^{O(k/\epsilon)} \cdot \poly(n, m)$,
where $n$ and $m$ are the numbers of vertices and edges in the input graph, respectively.
Thus, for fixed $\epsilon$, the algorithm runs in exponential time in $k$, while in polynomial time when $k$ is a fixed integer.

For the special case of $k = 2$, we obtain a $7/4$-approximation, improving the factor $2$ inherited from the MST-based $k$-approximation.
The algorithm combines bounded enumeration of cuts in a minimum spanning tree with contraction and spanning-tree packing.
When a suitable balanced edge does not exist, it uses a sink of an oriented minimum spanning tree and a solution to a constructed instance of the quota Steiner tree problem
to identify a low-cost portion that can be separated from the residual tree.

For {\sc BS$k$F} on metric complete graphs, we adapt the light-heavy component and matching viewpoint from min-max tree cover to the spanning-forest setting.
A vertex-disjoint tree-splitting argument preserves the partition constraint and leads to a polynomial-time $3$-approximation for arbitrary $k$,
improving the factor $4$ implied by Arkin et al.~\cite{AHL06}.

On the negative side, we prove the APX-hardness for {\sc BS$k$F} on metric complete graphs and on a special graph class,
by showing that in both cases, the problem cannot be approximated within $1688/1687 \approx 1.00059$ unless P $=$ NP.
The reductions start with the minimum vertex cover problem on graphs of maximum degree $5$, which was shown APX-hard by Chleb\'ik and Chleb\'ikov\'a~\cite{CC06}.
In particular, we use the degree-$5$ graph constructed by Chleb\'ik and Chleb\'ikov\'a from an instance of maximum linear equations using the consistency gadgets.

The remainder of the paper is organized as follows.
Section~\ref{sec02} introduces the notation and develops the structural tools used throughout the paper.
Section~\ref{sec:general-graphs} presents the approximation algorithms for general graphs, first for an arbitrary number of trees and then for two trees.
Section~\ref{sec:metric-complete} focuses on metric complete graphs and presents the constant ratio $3$-approximation algorithm.
The inapproximability results are in Section~\ref{sec:computational-complexity}, for metric complete graphs and for a sparse graph class.
We conclude the paper in Section~\ref{sec:conclusion} and discuss directions for further research.

\section{Preliminaries and structural tools}
\label{sec02}
In this section, we introduce the notations to be used throughout the paper,
formally describe the problems, and explain the structural tools to be employed in our approximation algorithms.

All graphs considered in the paper are finite and undirected.

We fix a connected graph $G = (V, E, w)$ for discussion, in which $V$ and $E$ are the vertex set and the edge set,
and $w: E \rightarrow \mbQ_{\ge 0}$ is a nonnegative edge weight function;
let $n = |V|$ and $m = |E|$.
For a positive integer $k$, we denote $[k] = \{1, \dots, k\}$.
A subgraph $H$ of $G$ itself is a graph, with the vertex set $V(H) \subseteq V$ and the edge set $E(H) \subseteq E$;
$H$ is a spanning subgraph of $G$ if $V(H) = V$.
For convenience, the resultant subgraph by removing a vertex $v$ and its incident edges from $G$ is denoted as $G - v$,
and the resultant subgraph by removing an edge $e$ from $G$ is denoted as $G - e$.

For an edge subset $F \subseteq E$, we extend the edge weight function to $F$ to denote $w(F) = \sum_{e \in F} w(e)$, and for a subgraph $H$ we abuse a bit by $w(H) = w(E(H))$.
This way, a single-vertex tree (also called a {\em singleton}) has weight $0$.
For a vertex subset $S \subseteq V$, $G[S]$ denotes the subgraph of $G$ induced by $S$, i.e., $V(G[S]) = S$ and $E(G[S])$ contains all the edges of $E$ with the two endpoints both in $S$.
If $G[S]$ is connected, $\MST(G[S])$ denotes a minimum spanning tree of $G[S]$.
Let $\delta_G(S)$ denote the subset of edges of $G$ with exactly one endpoint in $S$, i.e., $E(S, V \setminus S)$, called the {\em edge boundary} of $S$.
The degree of a vertex $v$ in $G$ is $\deg_G(v) = |\delta_G(\{v\})|$.

A {\em spanning $k$-forest} of $G$ is a spanning subgraph that is a collection $\mcT = \{T_1, \dots, T_k\}$ of exactly $k$ pairwise vertex-disjoint trees.
We study the following min-max optimization problem.

\begin{problem}[Balanced Spanning $k$-Forest, {\sc BS$k$F}]
\label{prob01}
~\\
{\bf Input:} A connected edge nonnegatively weighted graph $G = (V, E, w)$ and an integer $k$ with $2 \le k \le n$.\\
{\bf Objective:} Find a spanning $k$-forest $\mcT = \{T_1, \dots, T_k\}$ of $G$ minimizing $\max_{i\in[k]} w(T_i)$.
\end{problem}

In the above formulation of Problem~\ref{prob01}, the parameter $k$ is part of the input.
For a presented algorithm, when we analyze the running time, the exponent could depend only on $k$ and thus the algorithm is exponential time in $k$,
while it reduces to polynomial when $k$ is a fixed integer.
Also, given our min-max goal, we are aiming for exactly $k$ pairwise vertex-disjoint trees in any feasible solutions,
that is, a forest of less than $k$ pairwise vertex-disjoint trees can always be cut into exactly $k$ pairwise vertex-disjoint trees,
by removing some edges, without increasing the objective function value.

When there is no extra constraint on the input graph $G$, we call the {\sc BS$k$F} problem on {\em general} graphs.
We will consider also two restricted graph classes.
Let $\Delta(G)$ denote the maximum degree of $G$; $G$ is {\em subcubic} if $\Delta(G) \le 3$.
Graphs of {\em bounded degree} are those graphs of which $\Delta$ is upper bounded by a constant.
An edge weighted {\em complete} graph $G$ is called {\em metric} if the edge weights satisfy triangle inequality, i.e., $w(uv) \le w(uw) + w(wv)$ for any triple $u, v, w \in V$.

We also fix an optimal spanning $k$-forest $\mcT^* = \{T_1^*, \dots, T_k^*\}$ for the graph $G = (V, E, w)$ for discussion, and denote $\OPT = \max_{i \in [k]} w(T_i^*)$ the optimum.
Let $V_i^* = V(T_i^*)$ for each $i \in [k]$;
since $G[V_i^*]$ is connected, we assume without loss of generality that $T_i^* = \MST(G[V_i^*])$, i.e., $T_i^*$ is a minimum spanning tree of $G[V_i^*]$.

The next lemma states that {\sc BS$k$F} on trees can be solved exactly in $O(k n^3)$ time,
by an algorithm due to Vaishali et al.~\cite{VAP18} which is used as subroutine in our algorithms to solve such a boundary case.

\begin{lemma}{\rm\cite{VAP18}}
\label{lemma01}
The {\sc BS$k$F} problem on trees can be solved exactly in $O(k n^3)$ time.
\end{lemma}

Given a tree and a non-empty subset of its edges,
one can find the unique partition of the vertex set into two parts such that an edge belongs to this edge subset if and only if it connects exactly one vertex from each part.
The partition can be constructed in linear time by, first deleting the subset of edges from the given tree resulting in a collection of subtrees,
then merging the vertices of those subtrees that are at even distances into a part.
We define the following operation that generates such a bipartition.

\begin{operation}[Bipartition$(T, D)$]
\label{op01}
Given a tree $T$ and a nonempty edge subset $D \subseteq E(T)$,
the operation {\em Bipartition$(T, D)$} generates the unique partition $\{U_1, U_2\}$ of $V(T)$ such that $D = \delta_T(U_1) = \delta_T(U_2)$.
\end{operation}

\begin{lemma}
\label{lemma02}
Given a tree $T$ and a non-empty proper subset $S \subset V(T)$, let $D = \delta_T(S)$, then Bipartition$(T, D)$ generates the bipartition $\{S, V(T)\setminus S\}$.
\end{lemma}
\begin{proof}
Since $S$ is non-empty and proper, the edge subset $D = \delta_T(S)$ is non-empty.
The uniqueness of the partition generated by Bipartition$(T, D)$ guarantees that the partition is $\{S, V(T)\setminus S\}$.
\end{proof}

The next lemma on spanning-tree packing is due to Nash-Williams\cite{Nas61} and Tutte~\cite{Tut61}, and will be used repeatedly in this paper.
Note that in a multigraph, parallel edges connecting the same pair of vertices are regarded as distinct edges.

\begin{lemma}{\rm\cite{Nas61,Tut61}}
\label{lemma03}
Let $H$ be a multigraph and $b$ be a positive integer.
The graph $H$ contains $b$ pairwise edge-disjoint spanning trees if and only if, for every partition of $V(H)$ into $t \ge  2$ nonempty parts,
the number of edges of $H$ with endpoints in distinct parts is at least $b(t-1)$.
\end{lemma}

\begin{corollary}
\label{coro01}
Every $2b$-edge-connected multigraph contains $b$ pairwise edge-disjoint spanning trees.
\end{corollary}
\begin{proof}
Consider an arbitrary partition of the vertex set into $t \ge 2$ nonempty parts.
Since the graph is $2b$-edge-connected, the edge boundary of each part has size at least $2b$.
Summing these boundary sizes counts every edge connecting two distinct parts exactly twice, and thus the total number of such edges is at least $b t > b(t-1)$.
The conclusion then follows from \Cref{lemma03}.
\end{proof}

Let $T$ be a minimum spanning tree of the graph $G = (V, E, w)$.
For analysis only, we map the optimal spanning $k$-forest $\mcT^* = \{T_1^*, \dots, T_k^*\}$ to $T$ to construct a multigraph $G^*$ as follows:
We first identify all vertices of each $V_i^*$ into a single vertex denoted as $i$,
then delete the resulting loops (those edges of $T$ connecting two vertices of $V_i^*$ for some $i$),
and retain all parallel edges of $T$ connecting distinct sets $V_i^*$.
The resulting edge weighted multigraph is $G^*$ on $[k]$,
in which every retained edge inherits its weight from the corresponding edge of $T$ and they are all regarded distinct to each other.
For a subset $X \subseteq [k]$, let $V_X^* = \bigcup_{i\in X} V_i^*$.
By construction, the edge set $\delta_{G^*}(X)$ is the same as $\delta_T(V_X^*)$, though in different graphs.

\begin{lemma}
\label{lemma04}
Let $T$ be a minimum spanning tree of the graph $G = (V, E, w)$, and $G^*$ be the multigraph constructed above using the optimal spanning $k$-forest $\mcT^*$.
If $G^*$ contains $b$ pairwise edge-disjoint spanning trees for some integer $b \ge 2$, then $w(T) \le \frac b{b-1} k \cdot \OPT$.
\end{lemma}
\begin{proof}
The edge sets of these $b$ spanning trees in $G^*$ are pairwise disjoint and their union is a subset of edges of $T$.
It follows that one of these $b$ spanning trees has weight at most $w(T)/b$.
The edges in this tree spanning $[k]$ together with the sub-trees $T_1^*, \dots, T_k^*$ in $\mcT^*$ is a spanning tree of $G$.
Therefore, by the minimality of $T$, we have $w(T) \le \sum_{i\in[k]} w(T_i^*) + w(T)/b \le k \cdot \OPT + w(T)/b$, which can be rearranged into $w(T) \le \frac b{b-1} k \cdot \OPT$.
\end{proof}

Given a spanning tree $T$ of the graph $G = (V, E, w)$ and a vertex $v$,
some pairs of the connected components (which are trees) of $T - v$ could be connected by edges of $E \setminus E(T)$,
and we can construct an auxiliary graph on the connected components of $T - v$ through the following operation.

\begin{operation}[Auxiliary$(T, v)$]
\label{op02}
Given a spanning tree $T$ of the graph $G = (V, E, w)$ and a vertex $v \in V$,
let $C_1, \dots, C_d$ be the connected components of $T - v$, where $d = \deg_T(v)$, and $e_i$ be the unique edge of $T$ connecting $v$ to (a vertex of) $C_i$.
\begin{itemize}
\parskip=0pt
\item
	For each $C_i$, define its {\em profit} $p(C_i) = w(C_i) + w(e_i)$.
\item
	For two distinct $C_i$ and $C_j$, add an {\em auxiliary edge} $C_i C_j$ if $G$ contains an edge connecting $C_i$ and $C_j$,
	and the {\em cost} of the auxiliary edge is $c(C_i C_j) = \min\{w(xy): x \in V(C_i), y \in V(C_j), xy \in E\}$ and
	an original edge $xy$ attaining this minimum weight is stored with $C_i C_j$.
\end{itemize}
This constructs an {\em auxiliary graph} $H_v$ on the connected components of $T - v$ called {\em auxiliary vertices}, i.e., $\{C_1, \dots, C_d\}$,
with vertex-profit function $p$ and edge-cost function $c$.
\end{operation}

One sees that an auxiliary graph $H_v$ might not be connected.
For a vertex subset $X$ of $H_v$, we extend the profit function to denote $p(X) = \sum_{C \in X} p(C)$;
for any subgraph $K$ of $H_v$, we write $p(K) = p(V(K))$ and $c(K) = \sum_{e \in E(K)} c(e)$.
This way, $p(H_v) = w(T)$.

For each tree $T^*_j \in \mcT^*$ not containing the vertex $v$, those components of $T - v$ intersecting $T^*_j$ are connected in $H_v$,
and the cost of the minimum tree spanning these components is no greater than $w(T^*_j)$.
We summarize this and some more properties in the next lemma, which provides low-cost structures to be used in later constructions.

\begin{lemma}
\label{lemma05}
Given a minimum spanning tree $T$ of the graph $G = (V, E, w)$, the optimal spanning $k$-forest $\mcT^*$, and a vertex $v \in V^*_{i_0}$ for some $i_0 \in [k]$,
let $C_1, \dots, C_d$ be the components of $T - v$ and $H_v$ be the auxiliary graph returned by Operation~\ref{op02}.
For every $j \in [k] \setminus \{i_0\}$, let $\mcC_j= \{C_i: V(C_i) \cap V_j^* \ne \emptyset\}$.
Then $H_v$ contains a tree (called {\em auxiliary tree}) $Q_j$ on $\mcC_j$ with its cost $c(Q_j) \le w(T_j^*) \le \OPT$.
Moreover, the union of the components of $T - v$ spanned by these auxiliary trees has total profit $p\left(\bigcup_{j \in [k]\setminus\{i_0\}} \mcC_j \right) \ge w(T) - \OPT$.
\end{lemma}
\begin{proof}
Consider an index $j \ne i_0$.
One sees that the edge set $E(T_j^*)$ interconnects $\mcC_j$, and thus there is a minimum-cost spanning tree $Q_j$ on $\mcC_j$ in $H_v$.
The cost of $Q_j$ is no greater than the cost of a minimum-cost spanning tree if only the edges of $E(T_j^*)$ are used to define the edge-cost function $c$,
and therefore $c(Q_j) \le w(T^*_j) \le \OPT$.

Next, for each $C_i \notin \bigcup_{j \in [k]\setminus\{i_0\}} \mcC_j$, $V(C_i) \subseteq V^*_{i_0}$.
It follows that removing all these $C_i$'s and the edges connecting them to $v$ from $T$, followed by adding the tree $T^*_{i_0}$ to $T$,
give rise to a spanning connected subgraph of $G$, and thus its weight is at least as large as $T$ due to the minimality of $T$.
That is,
\[
p\left(\bigcup_{j \in [k]\setminus\{i_0\}} \mcC_j \right) + w(T^*_{i_0}) \ge w(T),
\]
which can be rearranged to prove the last statement.
\end{proof}

In our algorithms, we often want to select a connected set of auxiliary vertices with sufficiently large total profit at a low cost in a certain auxiliary graph,
which is exactly the {\em Quota Steiner Tree} ({\sc QST}) problem defined below.

\begin{problem}[Quota Steiner Tree, {\sc QST}]
\label{prob02}
~\\
{\bf Input:} A connected graph $H = (U, E_H, p, c)$, where $p$ is a nonnegative vertex-profit function and $c$ is a nonnegative edge-cost function, and a quota $L \in \mbQ_{> 0}$.\\
{\bf Objective:} Find a tree $R$ in $H$ such that $p(R) = \sum_{u \in V(R)} p(u) \ge L$ and $c(R) = \sum_{e \in E(R)} c(e)$ is minimized.
\end{problem}

One first sees that if $L > p(U)$ then the instance is infeasible, otherwise {\em feasible}.
Furthermore, when $H$ is disconnected, if at least one connected component is feasible, then we say the instance is feasible.
Also, when the vertex-profit function $p$ is uniform, that is, every vertex has the same profit, {\sc QST} reduces to the $k$-Minimum Spanning Tree ({\sc $k$-MST}) problem~\cite{JMP00}.
Johnson et al.~\cite{JMP00} show that an $\alpha$-approximation algorithm for {\sc $k$-MST} implies an $\alpha$-approximation algorithm for {\sc QST},
and Garg~\cite{Gar05} presents a $2$-approximation for $k$-MST.
Later we will call such an existing $2$-approximation algorithm for {\sc QST}, stated in the next lemma,
as a black-box subroutine for computing a desired connected set of auxiliary vertices.

\begin{lemma}[$2$-Approx for {\sc QST}~\cite{JMP00,Gar05}]
\label{lemma06}
Let $\OPT_{\mathrm{QST}}$ be the optimal cost of a feasible instance of the {\sc QST} problem.
There is a polynomial-time algorithm that returns a tree $R$ such that $p(R) \ge L$ and $c(R) \le 2 \cdot \OPT_{\mathrm{QST}}$.
\end{lemma}

We remark that in \Cref{prob02}, if the input graph $H$ is disconnected,
then one can simply call the $2$-approximation algorithm in \Cref{lemma06} on each feasible connected component of $H$ and retain the minimum-cost returned tree.

\begin{definition}[$\beta$-balanced edge]
\label{def01}
Let $0.5 < \beta < 1$.
Given a tree $T$ with nonnegative edge weights and an edge $e \in E(T)$, let $C_1^e$ and $C_2^e$ be the two connected components (which are trees) of $T - e$;
the edge $e$ is called {\em $\beta$-balanced} if $\max\{w(C_1^e), w(C_2^e)\} \le \beta w(T)$.
\end{definition}

\begin{lemma}
\label{lemma07}
Let $T$ be an edge nonnegatively weighted tree without any $\beta$-balanced edge, where $0.5 < \beta < 1$.
For every edge $e \in E(T)$, orient $e$ toward the heavier component of $T - e$ (whose weight is greater than $\beta w(T)$).
This orientation is well defined and has a sink $v$.
In the auxiliary graph $H_v$ returned by Operation~\ref{op02} using the induced subgraph $G[V(T)]$, $p(C_i) < (1 - \beta) w(T)$ for every vertex $C_i$ of $H_v$.
\end{lemma}
\begin{proof}
Because $e$ is not $\beta$-balanced, the heavier component of $T - e$ has weight greater than $\beta w(T)$ while the other less than $(1 - \beta) w(T) - w(e)$.
Thus the orientation is well defined.

The non-existence of a sink implies a directed cycle in the tree, a contradiction.

Lastly, a vertex $C_i$ of $H_v$ is the lighter component of $T - e$, where $e$ is the edge connecting $C_i$ and $v$,
and thus $w(C_i) < (1 - \beta) w(T) - w(e)$, which can be rearranged into $p(C_i) = w(C_i) + w(e) < (1 - \beta) w(T)$.
\end{proof}

\section{Approximation algorithms on general graphs}
\label{sec:general-graphs}
We present two approximation algorithms for {\sc BS$k$F} on general graphs, one for general $k \ge 2$ and the other for $k = 2$.

The next lemma states a $k$-approximation algorithm based on the minimum spanning tree of the input graph,
by which the solution is used as a candidate in our algorithms.

\begin{lemma}{\rm\cite{DMY25}}
\label{lemma08}
The {\sc BS$k$F} problem admits an $O(m \log n)$-time $k$-approximation algorithm, where $n$ and $m$ are the numbers of vertices and edges in the input graph.
\end{lemma}
\begin{proof}
Let $G = (V, E, w)$ be the input graph and $T$ be a minimum spanning tree of $G$, which can be computed in $O(m \log n)$-time.
Let $D$ be the set of the $k-1$ heaviest edges of $T$, then one sees that deleting $D$ from $T$ gives a spanning $k$-forest $\mcT$.
Let $D^* \subseteq E(T)$ be a set of $k-1$ edges interconnecting the $k$ vertex sets $V^*_1, \dots, V^*_k$ defined by the optimal solution $\mcT^* = \{T_1^*, \dots T_k^*\}$.
That is, the subgraph consisting of all the trees of $\mcT^*$ and the edges of $D^*$ is connected and thus it contains a spanning tree of $G$.
By the minimality of $T$, $w(T) \le \sum_{i \in [k]} w(T_i^*) + w(D^*) \le k \cdot \OPT + w(D^*) \le k \cdot \OPT + w(D)$.
It follows that every component of $T - D$ has weight at most $w(T) - w(D) \le k \cdot \OPT$.
\end{proof}

The following lemma explains how a $q$-forest in the auxiliary graph is expanded back to a spanning $(q+1)$-forest in the original graph.
Such a process will be employed in our algorithms.

\begin{lemma}
\label{lemma09}
Given a spanning tree $T$ of the graph $G = (V, E, w)$, a vertex $v \in V$, and the auxiliary graph $H_v$ on $\{C_1, \dots, C_d\}$ returned by Operation~\ref{op02},
let $\mcK = \{K_1, \dots, K_q\}$ be a $q$-forest in $H_v$.
For each $j \in [q]$, expand $K_j$ by replacing every auxiliary vertex $C_i \in V(K_j)$ with the corresponding component $C_i$ in $T - v$
and every auxiliary edge with its stored original edge, and denote the resulting subgraph of $G$ by $T_j$.
Let $T_0$ consist of $v$, all components $C_i$ in $T - v$ whose corresponding auxiliary vertices do not belong to $V(\mcK)$, and the corresponding tree edges $e_i$ connecting $C_i$ to $v$.
Then $\{T_0, T_1, \dots, T_q\}$ is a spanning $(q+1)$-forest of $G$, $w(T_j) \le p(K_j) + c(K_j)$ for every $j \in [q]$, and $w(T_0) = w(T) - p(\mcK)$.
\end{lemma}
\begin{proof}
For each tree $K_j \in \mcK$, since every $C_i \in V(K_j)$ is a tree in $T - v$ and the stored original edge corresponding to every edge of $K_j$ connecting two distinct components $C_i$.
The resultant $T_j$ is a tree in $G$.

Different trees of $\mcK$ are on disjoint sets of auxiliary vertices, so the trees $T_1, \dots, T_q$ are pairwise vertex-disjoint.
The subgraph $T_0$ is a subtree of $T$ containing $v$ and the components of $T - v$ not included in any $T_j$ for $j \in [q]$, and therefore is a tree.
This proves that $\{T_0, T_1, \dots, T_q\}$ is a spanning $(q+1)$-forest in $G$.

For each $j \in [q]$ and every $C_i \in V(K_j)$, the edge $e_i$ in $T$ connecting the component $C_i$ to $v$ is excluded from $T_j$,
and thus the total weight of components in $V(K_j)$ is no greater than $p(K_j)$;
the stored original edges included in $T_j$ have total weight $c(K_j)$.
Hence $w(T_j) \le p(K_j) + c(K_j)$.
Lastly, since the sets $E(C_i) \cup \{e_i\}$, where $e_i$ connects $C_i$ to $v$ in $T$, partition $E(T)$, we have $w(T_0) = w(T) - p(\mcK)$.
\end{proof}

\subsection{$k \ge 2$}
\label{subsec:general-bsf-k}
There are two main design ideas in our algorithm, both working with a minimum spanning tree $T$ of the input graph $G = (V, E, w)$.
Firstly, for an edge set $D \subseteq E(T)$ of small size
(which is determined by the spanning tree {\em packing} parameter $b$ introduced in \Cref{lemma04}, that controls the breadth of the recursion),
the operation Bipartition$(T, D)$ returns two sets $U_1$ and $U_2 = V \setminus U_1$;
we recursively compute a spanning $q$-forest of $G[U_1]$ and a spanning $(k - q)$-forest of $G[U_2]$,
and then merge them into a spanning $k$-forest for the original graph $G$.
Secondly, at the presence of a $\beta$-balanced edge $e \in E(T)$,
$T - e$ is a very good starting point with suitable $\beta$ and further removing $k - 2$ other edges from $T$ gives a good candidate $k$-forest of $G$;
in the other case, the existence of a sink $v$ by \Cref{lemma07} and then
the auxiliary graph $H_v$ constructed by the operation Auxiliary$(T, v)$ presents a good structure for constructing a $(q-1)$-forest,
which can be expanded back to a spanning $q$-forest, and subsequently a candidate $k$-forest if $q < k$, in the original graph $G$.

The two main design ideas interweave.
The second case does not get into recursion but every recursion has it to construct some candidate solutions;
nevertheless, it can report ``{\em failure}'' if no desired solution is found.
There are two basic boundary scenarios where an optimal solution is returned:
In the first scenario where the recursion calls for a spanning $1$-forest, a minimum spanning tree is optimal.
In the second scenario where a minimum spanning tree containing at most $q-1$ edges of positive weight is found, a spanning $q$-forest of weight $0$ is optimal.
We therefore assume without loss of generality in the following discussion that
our target is a spanning $q$-forest with $2 \le q \le k$ and the current minimum spanning tree $T$ has weight $w(T) > 0$.

We first present a subroutine {\sc QST}-Forest that calls the $2$-approximation algorithm for the {\sc QST} problem
to compute an auxiliary forest with a bounded number of trees in a given auxiliary graph $H$.
Importantly, the subroutine, with its description in \Cref{alg01}, might not succeed in finding a desired solution and, in this case, it returns {\em failure}.

\begin{algorithm}[htb]
\caption{{\sc QST}-Forest $((H, p, c), L, q)$\label{alg01}}
\begin{algorithmic}[1]
\Require
	An auxiliary graph $H = (U, E, p, c)$, a quota $L > 0$, and an integer $q \ge 1$;
\Ensure
	An auxiliary forest $\mcK$ with at most $q$ trees and $p(\mcK) \ge L$, or {\em failure} if no such forest is found.
\ForAll {($Z \subseteq U$ with $1 \le |Z| \le q$)}
	\State Add a virtual root $z$ with $p(z) = 0$;
	\State add a zero-cost edge $zu$ for every $u \in Z$;
	\If {($(H_Z, p, c, L)$ is feasible)}\Comment{$H_Z$ denotes the augmented graph;}
		\State call the $2$-approximation of \Cref{lemma06} with quota $L$;
		\State delete the virtual root and virtual edges from the returned tree $R$;
		\State record the resulting forest if no forest has been recorded yet or if its cost is smaller;
	\EndIf
	\State remove the virtual root and virtual edges;
\EndFor
\State \Return the recorded forest, or failure if none was recorded.
\end{algorithmic}
\end{algorithm}

\begin{lemma}
\label{lemma10}
Given an auxiliary graph $H = (U, E, p, c)$, a quota $L > 0$, and an integer $q \ge 1$,
if $H$ contains an auxiliary forest $\mcK_0$ with at most $q$ trees and $p(\mcK_0) \ge L$,
then Algorithm~\ref{alg01} returns an auxiliary forest $\mcK$ with at most $q$ trees, $p(\mcK) \ge L$, and $c(\mcK) \le 2c(\mcK_0)$.
The running time of the algorithm is $|U|^{O(q)} \poly(|U|,|E|)$.
\end{lemma}
\begin{proof}
One sees that by choosing one vertex from each tree of $\mcK_0$ to form $Z$, the resulting revised $H$ is a feasible {\sc QST} instance with $\OPT \le c(\mcK_0)$.
The $2$-approximation of \Cref{lemma06} thus returns a tree $R$ with $p(R) \ge L$ and $c(R) \le 2 \cdot \OPT$;
the resulting forest $\mcK$ from $R$ contains at most $q$ trees, $p(\mcK) = p(R) \ge L$, and $c(\mcK) = c(R) \le 2 c(\mcK_0)$.
The final returned forest by \Cref{alg01} is feasible, and with cost not larger than $c(\mcK)$.

\Cref{alg01} iterates through all subsets of size up to $q$, of which there are $|U|^{O(q)}$,
and each iteration calls the polynomial-time $2$-approximation for {\sc QST}.
The overall running time is $|U|^{O(q)} \poly(|U|,|E|)$.
\end{proof}

The next lemma controls the profit of an auxiliary forest in an auxiliary graph by repeatedly pruning away a leaf or an isolated vertex,
without increasing its cost or its number of trees.

\begin{lemma}
\label{lemma11}
Given an auxiliary forest $\mcK$ with $p(\mcK) \ge L' > 0$ in an auxiliary graph $H$ where every auxiliary vertex $C$ satisfies $p(C) < L$,
repeatedly delete a leaf vertex or an isolated vertex $C$ from $\mcK$ whenever $p(\mcK - C) \ge L'$.
When no such deletion is possible, the resulting forest satisfies $L' \le p(\mcK) < L' + L$.
The pruning does not increase the cost of $\mcK$ or the number of trees in $\mcK$.
\end{lemma}
\begin{proof}
One sees that every deletion preserves $\mcK$ being a forest, preserves $p(\mcK) \ge L'$, decreases the number of vertices by one, possibly decreases the number of edges by one,
and possibly decreases the number of trees by one.
If there is only one vertex remained in $\mcK$, then $L' \le p(\mcK) < L < L' + L$.
Otherwise, $\mcK$ contains at least two leaves or at least two isolated vertices, and $p(\mcK) \ge L' + L$ implies the lighter leaf or isolated vertex can be deleted.
This proves the lemma.
\end{proof}

We now use the above two lemmas to construct a spanning $q$-forest for the graph $G = (V, E, w)$, where $3 \le q \le k$, from an auxiliary graph $H_v$ where $v$ is a sink.
In this case, by \Cref{lemma07}, the tree $T$ leading to $H_v$ has no $\beta$-balanced edge, where $0.5 < \beta < 1$, and $p(C) < (1 - \beta) w(T)$ for every auxiliary vertex $C$ of $H_v$.

A description of the construction process is in \Cref{alg02}, for which the input is the graph $G = (V, E, w)$,
a minimum spanning tree $T$ containing no $\beta$-balanced edge, an integer $3 \le q \le \min\{k, |V|\}$, an integer $1 \le s \le q-1$, and a quota $s L > 0$.
The process starts with adding the single-vertex trees into the auxiliary forest $\mcK$ whenever $p(\mcK) < s L$ and the auxiliary vertex has profit greater than or equal to $L$.
During this adding phase, if $p(\mcK)$ reaches $s L$, then adding stops and $\mcK$ is expanded by \Cref{lemma09} into a spanning forest $\mcT$ of $G$ with no more than $s + 1 \le q$ trees,
and $\mcT$ can be further converted into a spanning $q$-forest by deleting some edges.
Otherwise, at the end of adding phase $p(\mcK) < s L$ and every auxiliary vertex has profit less than $L$.
The process calls \Cref{alg01} on the remainder auxiliary graph with the updated quota $L'$ and the updated bound $r$ on the number of trees
to compute another forest $\mcK'$ of $H_v$ with at most $r$ trees, which is subsequently pruned by \Cref{lemma11} to have profit in between $L'$ and $L' + L$.
Appending $\mcK'$ to $\mcK$ gives rise a forest with at most $s$ trees, which goes through the expanding process to achieve a spanning $q$-forest of $G$.
Again importantly, \Cref{alg01} might fail in returning a desired forest and, in this case, the construction process also returns {\em failure}.

\begin{algorithm}[htb]
\caption{Auxiliary-$q$-Forest $(G = (V, E, w), T, \beta, q, s, L)$\label{alg02}}
\begin{algorithmic}[1]
\Require
	A connected graph $G = (V, E, w)$, a minimum spanning tree $T$ containing no $\beta$-balanced edge, where $0.5 < \beta < 1$,
	an integer $3 \le q \le \min\{k, |V|\}$, an integer $1 \le s \le q-1$, and a quota $s L > 0$;
\Ensure
	A spanning $q$-forest of $G$, or {\em failure} if no such forest is found.
\State Orient the edges of $T$ as in \Cref{lemma07}, choose a sink $v$, and construct $H_v$ by Operation~\ref{op02};
\State initialize $\mcK$ as an empty auxiliary forest;
\While {($p(\mcK) < s L$, $\exists C \notin V(\mcK)$ with $p(C) \ge L$)}
	\State add the single-vertex tree $\{C\}$ to $\mcK$;
	\If {($p(\mcK) \ge sL$)}
		\State proceed to Line~15;\Comment{$\mcK$ contains at most $s$ trees;}
	\EndIf
\EndWhile
\State set $L' \gets s L - p(\mcK)$;\Comment{$\mcK$ contains less than $s$ trees, $p(\mcK) < s L$, and $p(C) < L, \forall C \notin V(\mcK)$;}
\State define $p'$ by $p'(C) = 0$ for $C \in V(\mcK)$ and $p'(C) = p(C)$ otherwise;
\State set $r \gets \left\lceil \frac {(q-1)L'}{L'+(q-1)L} \right\rceil$ and call \Cref{alg01} on $((H_v, p', c), L', r)$ for a forest $\mcK'$;
\If {($\mcK'$ failed)}
	\State\Return failure;
\Else
	\State prune $\mcK'$ by \Cref{lemma11} using the profit function $p'$;
	\State set $\mcK \gets \mcK' \cup \mcK$ followed by cycle breaking;\Comment{$\mcK$ contains at most $s$ trees;}
\EndIf
\State expand $\mcK$ by \Cref{lemma09} to obtain a spanning forest $\mcT$ of $G$;\Comment{$\mcT$ contains at most $s+1 \le q$ trees;}
\State repeatedly delete the heaviest edge of $\mcT$ until $\mcT$ becomes $q$-forest;
\State\Return $\mcT$.
\end{algorithmic}
\end{algorithm}

\begin{lemma}
\label{lemma12}
If the process in \Cref{alg02} returns a forest $\mcT$ of the graph $G$, then $\mcT$ is a spanning $q$-forest and
the predecessor $\mcK$ is an auxiliary forest of $H_v$ with $p(\mcK) \ge s L$ and at most $s$ trees.
\end{lemma}
\begin{proof}
Note that there are two ways of obtaining $\mcK$, followed by expanding it by \Cref{lemma09} into $\mcT$.
In the first way, $\mcK$ contains only single-vertex trees, each corresponding to an auxiliary vertex of $H_v$ with profit greater than or equal to $L$.
Since no more single-vertex tree can be added once $p(\mcK) \ge s L$, we conclude that $\mcK$ has at most $s$ such trees.

In the other case, suppose $\mcK$ has $p(\mcK) < s L$ and $t$ such single-vertex trees at the end of the adding phase.
All the other auxiliary vertices have their profit less than $L$.
By setting $L' = s L - p(\mcK) \le (s - t) L$ and $r = \left\lceil \frac {(q-1)L'}{L'+(q-1)L} \right\rceil \le s - t$,
the forest $\mcK'$ returned by \Cref{alg01} has $p'(\mcK') \ge L'$ and at most $r$ trees.
After pruning by \Cref{lemma11}, $L' \le p'(\mcK') < L' + L$ and $\mcK'$ still has at most $r$ trees.
It follows that unionizing $\mcK'$ to $\mcK$, followed by deleting an edge from each cycle, gives rise to $s L \le p(\mcK) < (s+1) L$ and at most $t + r \le s$ trees inside.

Expanding $\mcK$ by \Cref{lemma09} into a spanning forest $\mcT$ of $G$ increases the number of trees by $1$,
that is, $\mcT$ contains at most $s+1 \le q$ trees to start with, and can be converted into a $q$-forest by removing some edges if necessary.
This proves the lemma.
\end{proof}

We are now ready to present the main algorithm that combines recursion on bipartitions obtained by cutting a small set of edges,
and spanning $q$-forest direct constructions at the presence of a $\beta$-balanced edge (cf. \Cref{def01}) and at no presence of any $\beta$-balanced edge (by \Cref{alg02}).
The algorithmic description is in \Cref{alg03}, which returns a spanning $q$-forest of the input graph.
The spanning tree {\em packing} parameter $b \ge 5$ is a fixed constant used to control the breadth of the recursion, of which the value is determined later in the performance analysis.

\begin{algorithm}[htb]
\caption{General-$q$-Forest $(G = (V, E, w), q, b)$\label{alg03}}
\begin{algorithmic}[1]
\Require
	A connected graph $G = (V, E, w)$, an integer $1 \le q \le \min\{k, |V|\}$, an integer $b \ge 5$;
\Ensure
	A spanning $q$-forest of $G$.
\State Compute a minimum spanning tree $T$ of $G$;
\State convert $T$ into a spanning $q$-forest $\mcT$ by \Cref{lemma08}, and record $\mcT$;\Comment{A candidate is guaranteed;}
\If {($q = 1$ or $\max_{T \in \mcT} w(T) = 0$)}\Comment{Boundary scenarios, recursion ends;}
	\State\Return $\mcT$;
\EndIf
\ForAll {($D \subseteq E(T)$ with $1 \le |D| \le \min\{2b-1, |V|-1\}$)}\Comment{Recursive calls on small cuts;}
	\State $\{U_1, U_2\} = {\rm Bipartition}(T, D)$ by \Cref{op01};
	\If {(both $G[U_1]$ and $G[U_2]$ are connected)}
		\For {($s = 1$ to $q-1$)}
			\If {($|U_1| \ge s$ and $|U_2| \ge q-s$)}
				\State recursively solve $(G[U_1], s, b)$ and $(G[U_2], q-s, b)$;
				\State record the union of the returned spanning $s$-forest and spanning $(q-s)$-forest;
			\EndIf
		\EndFor
	\EndIf
\EndFor
\If {($q \ge 3$)}\Comment{Direct spanning $q$-forest constructions;}
	\For {($s = 1$ to $q-1$)}
		\State set $L \gets \frac{(1 - 2/q) w(T)}{s + q -1}$ and $\beta \gets 1 - \frac {sL}{w(T)}$;
		\If {($T$ contains a $\beta$-balanced edge)}
			\State delete such an edge and $q-2$ other edges from $T$, and record the resulting $q$-forest;
		\Else
			\State call \Cref{alg02} using $(G, T, \beta, q, s, L)$, and record the succeeded $q$-forest;
		\EndIf
	\EndFor
\EndIf
\State\Return the recorded $q$-forest of minimum objective value.
\end{algorithmic}
\end{algorithm}

\begin{theorem}
\label{thm01}
Given a connected graph $G = (V, E, w)$, an integer $1 \le q \le \min\{k, |V|\}$, and an integer $b \ge 5$, Algorithm~\ref{alg03} returns a spanning $q$-forest of $G$.
Its running time at the top-level call with $q = k$ is $n^{O(kb)} \poly(n, m)$, where $n = |V|$ and $m = |E|$.
\end{theorem}
\begin{proof}
We prove feasibility by induction on $q$.
For $q = 1$, the algorithm returns a minimum spanning tree.
Let $q \ge 2$ and assume that all calls for a forest with fewer trees return desired forests.
\Cref{lemma08} produces a spanning $q$-forest in the current call, so a candidate is always recorded.
In the for-loop of Algorithm~\ref{alg03} enumerating all small edge sets (Lines 5--11),
each recursive call returns successfully a forest with fewer than $q$ trees by the induction hypothesis, and each union of two such forests is a spanning $q$-forest of the current graph.

In the direct $q$-forest construction phase (Lines 12--18), $L > 0$ because $q \ge 3$ and $w(T) > 0$.
Moreover, since $s \le q-1$, $1 - \beta = \frac {s(1 - 2/q)}{s+q-1} < 0.5$, that is, $\beta > 0.5$.
At the existence of a $\beta$-balanced edge, a spanning $q$-forest is obtained by deleting $q-1$ edges from the minimum spanning tree;
\Cref{lemma12} guarantees a spanning $q$-forest whenever Algorithm~\ref{alg02} succeeds.
Therefore, every recorded forest is a spanning $q$-forest.

For the running time, the recursion phase enumerates $n^{O(b)}$ edge sets $D$ and at most $q-1$ values of $s$.
Its depth is thus at most $q-1$ because all recursive calls are for a forest with fewer trees.
Therefore, the total number of recursive calls at the top-level with $q = k$ is at most $n^{O(kb)} \poly(n, m)$.
In the direct $q$-forest construction phase, at most $q-1$ values of $s$ are considered.
In \Cref{alg02}, since $r \le s \le k-1$, \Cref{lemma10} bounds each call by $n^{O(k)} \poly(n, m)$, which is dominated by $n^{O(kb)} \poly(n, m)$.
This proves the theorem.
\end{proof}

We next analyze the quality of the returned spanning $q$-forest of the graph $G = (V, E, w)$ by Algorithm~\ref{alg03} on $(G = (V, E, w), q, b)$.
Let $T$ denote the minimum spanning tree of $G$, $\mcT^* = \{T^*_1, \dots, T^*_q\}$ denote an optimal spanning $q$-forest of $G$, $\OPT = \max_{i \in [q]} w(T^*_i)$,
and $G^*$ denote the contracted multigraph constructed using $T$ and $\mcT^*$ (right before \Cref{lemma04}).
The next lemma states that either the value of $\OPT$ is passed on to some pair of smaller instances recursively called on,
or the weight of $T$ is upper bounded by a constant multiple of $\OPT$.

\begin{lemma}
\label{lemma13}
If $G^*$ has an edge cut of size $2b-1$ or less,
then the for-loop of Algorithm~\ref{alg03} considers a bipartition $\{U_1, U_2\}$ that, for some pair $(s, q-s)$,
both the optimum value of spanning $s$-forest of $G[U_1]$ and the optimum value of spanning $(q-s)$-forest of $G[U_2]$ are at most $\OPT$.
Otherwise, $G^*$ is $2b$-edge-connected and $w(T) \le \frac b{b-1} q \cdot \OPT$.
\end{lemma}
\begin{proof}
If $G^*$ has an edge cut of size at most $2b-1$ then, since $G^*$ is connected, a minimum-size edge cut $\delta_{G^*}(X)$ has size at most $2b-1$ to separate $X$ from $V(G^*) \setminus X$,
so that $G^*[X]$ and $G^*[V(G^*) \setminus X]$ are connected.

Let $U_1 = V_X^*$ and $U_2 = V \setminus U_1$.
Since $\delta_{G^*}(X)$ and $D = \delta_T(U_1)$ are the same set of edges in different graphs, $D$ is enumerated by the for-loop of Algorithm~\ref{alg03},
and Lemma~\ref{lemma02} states that Operation~\ref{op01} returns the bipartition $\{U_1, U_2\}$.
Since $G^*[X]$ and $G^*[V(G^*) \setminus X]$ are connected in $G^*$ and each optimal tree $T^*_i$ is connected in $G$, both $G[U_1]$ and $G[U_2]$ are connected in $G$.
When $s = |X|$, the optimal trees $T^*_i$ of $X$ form a feasible spanning $s$-forest of $G[U_1]$, and the remaining optimal trees form a feasible spanning $(q-s)$-forest of $G[U_2]$.
Therefore, the optimum value of spanning $s$-forest of $G[U_1]$ and the optimum value of spanning $(q-s)$-forest of $G[U_2]$ are both at most $\OPT$.
This proves the first half of the lemma.

If no edge cut of size at most $2b-1$ exists, then $G^*$ is $2b$-edge-connected.
\Cref{coro01} guarantees $b$ pairwise edge-disjoint spanning trees in $G^*$, and then by \Cref{lemma04} we have $w(T) \le \frac b{b-1} q \cdot \OPT$.
This proves the second half of the lemma.
\end{proof}

For the direct spanning $q$-forest constructions, there are two branches on the existence of a $\beta$-balanced edge or not in the minimum spanning tree $T$.
We first examine the slightly more involved branch of non-existence in \Cref{lemma14}, and then bound the quality of the best recorded spanning $q$-forest in \Cref{lemma15}.

\begin{lemma}
\label{lemma14}
Suppose that $q \ge 3$, $w(T) > \frac q2 \cdot \OPT$,
and the algorithm considers a value $s \in \{1, \dots, q-1\}$ for which $T$ has no $\beta$-balanced edge with respect to the parameters $L$ and $\beta$.
Then Algorithm~\ref{alg02} on $(G = (V, E, w), T, \beta, q, s, L)$ does not fail.
The predecessor auxiliary forest $\mcK$ of the auxiliary graph $H_v$ has at most $s$ trees and $p(\mcK) \ge sL$.
If $\mcK$ breaks out of the while-loop, then $c(\mcK) = 0$;
otherwise, we have $p(\mcK) < (s+1)L$ and $c(\mcK) \le 2 \left\lceil \frac {s(q-1)}{s+q-1} \right\rceil \cdot \OPT$.
\end{lemma}
\begin{proof}
Note that the algorithm sets $L = \frac {1 - 2/q}{s+q-1} w(T)$ and $\beta = 1 - \frac {sL}{w(T)}$.
It follows that $(1 - \beta) w(T) = sL$.
The non-existence of a $\beta$-balanced edge, by \Cref{lemma07}, implies $p(C) < sL$ for every auxiliary vertex $C$ in $H_v$.

Assume the sink $v \in V_{i_0}^*$.
\Cref{lemma05} provides trees $Q_j$, for $j \in [q] \setminus \{i_0\}$,
each of cost at most $\OPT$ and their union of profit at least $w(T) - \OPT > (1 - 2/q) w(T) = (s+q-1)L$ due to $w(T) > \frac q2 \cdot \OPT$.

If $\mcK$ breaks out of the while-loop (Line 6), then \Cref{alg02} succeeds and
(see also \Cref{lemma12}) $\mcK$ consists only of isolated auxiliary vertices and hence $c(\mcK) = 0$.

Otherwise, each vertex not selected by $\mcK$ has profit less than $L$, and \Cref{alg02} sets $L' = sL - p(\mcK) > 0$ and
modifies the profit function $p'$ to remove at most $p(\mcK) = sL - L'$ from the union of the above trees $Q_j$.
That is, the union of these trees has $p'$-profit greater than $L' + (q-1)L$.
\Cref{alg02} further sets $r = \left\lceil \frac {(q-1)L'}{L'+(q-1)L} \right\rceil$ (Line 9) to call \Cref{alg01} on the created instance $((H_v, p', c), L', r)$.

We claim that \Cref{alg01} does not fail.
Choose uniformly at random $r$ of the $q-1$ trees $Q_j$.
Every auxiliary vertex covered by the union of all the $q-1$ trees is covered by the random subfamily of $r$ with probability at least $r/(q-1)$.
Therefore, the expected $p'$-profit of the covered auxiliary vertices is greater than $(L' + (q-1)L) \cdot r/(q-1) \ge L'$.
That is, some subfamily of $r$ covers $p'$-profit greater than $L'$ with cost at most $r \cdot \OPT$.
By \Cref{lemma10}, \Cref{alg01} succeeds and returns a forest $\mcK'$ of cost at most $2r \cdot \OPT$.

Because each vertex not selected by $\mcK$ has profit (and thus $p'$-profit) less than $L$ and each vertex in $\mcK$ has zero $p'$-profit,
\Cref{lemma11} successfully prunes $\mcK'$ to ensure that $L' \le p'(\mcK') < L' + L$.
Therefore, at the end (see also \Cref{lemma12}) the produced forest $\mcK$ has at most $s$ trees, $s L \le p(\mcK) < (s+1) L$,
and $c(\mcK) \le 2r \cdot \OPT \le 2 \left\lceil \frac {s(q-1)}{s+q-1} \right\rceil \cdot \OPT$.
This proves the lemma.
\end{proof}

\begin{lemma}
\label{lemma15}
Suppose that $q \ge 3$, $w(T) > \frac q2 \cdot \OPT$, and $G^*$ has no edge cut of size $2b-1$ or less.
Let $g = b/(b-1)$ and $\alpha_b = \frac {g(g+2)}{2(g+1)}$.
Algorithm~\ref{alg03} on $(G = (V, E, w), q, b)$ returns a spanning $q$-forest of $G$ with objective value at most $(\alpha_b q + 3) \OPT$.
\end{lemma}
\begin{proof}
Let $\theta = \frac g{g+2}$;
then $\alpha_b = \frac {g(g+2)}{2(g+1)} = \frac g{1 + \theta}$.

Since $G^*$ has no edge cut of size $2b-1$ or less, by \Cref{lemma13} we have $w(T) \le g q \cdot \OPT$.

Since $b \ge 5$, we have $1 < g = b/(b-1) \le 5/4$ and $1/3 < \theta = g/(g+2) \le 5/13$.
Using $q \ge 3$, we have $1 < \theta q < q - 1.8$.
Consider the iteration of the direct spanning $q$-forest constructions where $s = \lfloor \theta q \rfloor$ (satisfying $1 \le s \le q-2$).

In this iteration, with the parameters $L$ and $\beta$,
if the minimum spanning tree $T$ of $G$ has a $\beta$-balanced edge $e$,
then each tree in the recorded spanning $q$-forest resulted from deleting $e$ and other $q-2$ edges from $T$ has weight at most $\beta w(T)$.
Otherwise, \Cref{lemma14} guarantees an auxiliary forest $\mcK$ with $p(\mcK) \ge sL$,
and then by \Cref{lemma09} the residual tree $T_0$ after expansion has weight at most $w(T) - sL = \beta w(T)$.
If $\mcK$ breaks out of the while-loop, then it consists of single-vertex trees each of profit, by \Cref{lemma07}, at most $(1-\beta) w(T) < \beta w(T)$;
if it does not, then by \Cref{lemma14} $p(\mcK) < (s+1) L$ and $c(\mcK) \le 2 \left\lceil \frac {s(q-1)}{s+q-1} \right\rceil \cdot \OPT$,
and thus every tree in the expanded $q$-forest of $G$ has weight at most $p(\mcK) + c(\mcK) < (s+1) L + 2 \left\lceil \frac {s(q-1)}{s+q-1} \right\rceil \cdot \OPT$.
We bound these two quantities in the following.

Recall the definitions $L = \frac {(1 - 2/q) w(T)}{s + q -1}$ and $\beta = 1 - \frac {sL}{w(T)}$.
Using the above bound $w(T) \le g q \cdot \OPT$, we have
\[
\beta w(T) = \frac {q(q-1)+2s}{q(s+q-1)} w(T) \le \frac {g (q(q-1)+2s)}{s+q-1} \OPT,
\]
of which the last expression is strictly decreasing in $s$ due to $q \ge 3$ and $b \ge 5$.
Since $s = \lfloor \theta q \rfloor > \theta q-1$,
\[
\beta w(T) < \frac {g (q(q-1) + 2 \theta q)}{\theta q - 1 + q-1} \OPT = \left( \frac {gq}{1+\theta} + \frac {gq(2\theta^2 + \theta + 1)}{(1 + \theta)((1 + \theta)q - 2)} \right) \OPT,
\]
where the term $\frac {gq(2\theta^2 + \theta + 1)}{(1 + \theta)((1 + \theta)q - 2)}$ can be verified to be smaller than $387/160 < 3$ using $q \ge 3$ and $b \ge 5$.
We therefore prove that $\beta w(T) < (\alpha_b q + 3) \OPT$.

One the other hand, also using the above bound $w(T) \le g q \cdot \OPT$, we have
\[
\Gamma := (s+1) L + 2 \left\lceil \frac {s(q-1)}{s+q-1} \right\rceil \cdot \OPT \le \left( \frac {g(q-2)(s+1) + 2 s (q-1)}{s+q-1} + 2 \right) \OPT,
\]
of which the last expression is strictly increasing in $s$ due to $q \ge 3$ and $b \ge 5$.
Since $s = \lfloor \theta q \rfloor \le \theta q$,
\[
\Gamma \le \left( \frac {g(q-2)(\theta q + 1) + 2 \theta q (q-1)}{\theta q + q-1} + 2 \right) \OPT = \left( \frac {gq}{1+\theta} + 3
	- \frac {q(\theta^3+\theta^2-\theta+1)+5\theta^2+4\theta-1}{(1-\theta^2)((1+\theta)q-1)} \right) \OPT,
\]
where the term $\frac {q(\theta^3+\theta^2-\theta+1)+5\theta^2+4\theta-1}{(1-\theta^2)((1+\theta)q-1)}$ can be verified to be positive using $q \ge 3$ and $b \ge 5$.
We therefore prove that $(s+1) L + 2 \left\lceil \frac {s(q-1)}{s+q-1} \right\rceil \cdot \OPT < (\alpha_b q + 3) \OPT$.

This upper bound of $(\alpha_b q + 3) \OPT$ on the weight of every tree in the above two explicit recorded spanning $q$-forests proves the lemma.
\end{proof}

We are now ready to prove the approximation ratio of \Cref{alg03} in the next theorem.

\begin{theorem}
\label{thm02}
Given a connected graph $G = (V, E, w)$, an integer $k \ge 2$, and an integer $b \ge 5$,
Algorithm~\ref{alg03} returns a spanning $k$-forest of $G$ with objective value at most $(\alpha_b k + 3) \OPT$, where $\alpha_b = \frac {b(3b-2)}{2(b-1)(2b-1)}$.
\end{theorem}
\begin{proof}
We prove the approximation ratio by strong induction on the number $q$, where $1 \le q \le \min\{k, |V|\}$, of trees required by the current recursive call.
We have $\frac 34 < \alpha_b \le \frac {65}{72}$ as it is strictly decreasing in $b$.

For the base case $q = 1$, the algorithm returns a minimum spanning tree and is optimal.
For $q = 2$, \Cref{lemma08} gives a solution of value at most $2\,\OPT \le (2 \alpha_b + 3) \OPT$.
Assume in the current instance that $q \ge 3$ and that the claim holds for all values smaller than $q$.
Let $T$ be the minimum spanning tree of the graph $G$.
We distinguish three cases:

{\bf Case 1.} $w(T) \le \frac q2 \cdot \OPT$.
In this case, the spanning $q$-forest by \Cref{lemma08} has objective value at most $w(T) \le \frac q2 \cdot \OPT \le (\alpha_b q + 3) \OPT$.

Assume henceforth that $w(T) > \frac q2 \cdot \OPT$.

{\bf Case 2.} The contracted multigraph $G^*$ constructed using $T$ and the optimal spanning $q$-forest $\mcT^*$ (right before \Cref{lemma04}) has an edge cut of size $2b-1$ or less.
In this case, by \Cref{lemma13}, the for-loop of Algorithm~\ref{alg03} considers a bipartition $\{U_1, U_2\}$,
for some pair $(s, q-s)$, the optimum value of spanning $s$-forest of $G[U_1]$ and the optimum value of spanning $(q-s)$-forest of $G[U_2]$ are at most the current $\OPT$.
It follows that when this pair $(s, q-s)$ is used in the recursive calls,
by the inductive assumption, the returned forests have objective value at most $(\alpha_b s + 3) \OPT$ and $(\alpha_b (q-s) + 3) \OPT$, respectively,
both are at most $(\alpha_b q + 3) \OPT$.
That is, such a spanning $q$-forest satisfies the desired bound.

{\bf Case 3.} $G^*$ has no edge cut of size $2b-1$ or less.
In this case, \Cref{lemma15} guarantees a directly constructed spanning $q$-forest of objective value at most $(\alpha_b q + 3) \OPT$.

These three cases complete the induction, and the top-level value $q = k$ proves the theorem.
\end{proof}

\begin{corollary}
\label{coro02}
Let $0 < \epsilon < 0.25$, set $b = \left\lceil \frac 5{2\epsilon} \right\rceil + 1$, and let $k \ge \lceil 6/\epsilon \rceil$.
Then \Cref{alg03} is a $(0.75 + \epsilon) k$-approximation algorithm for the {\sc BS$k$F} problem,
and its running time is $n^{O(k/\epsilon)} \poly(n, m)$, which is polynomial for any fixed $k$ and is exponential parameterized by $k$ when $k$ is part of the input.
\end{corollary}
\begin{proof}
The coefficient in Theorem~\ref{thm02} is $\alpha_b = \frac 34 + \frac {5b-3}{4(b-1)(2b-1)}$.
Since $b-1 \ge \frac 5{2\epsilon}$, we have $\alpha_b - \frac 34 < \frac 5{4(b-1)} \le \frac {\epsilon}2$.
Also, $k \ge \lceil6/\epsilon\rceil$ implies $3 \le \epsilon k/2$.
Therefore, $\alpha_b k + 3 \le (\frac 34 + \frac {\epsilon}2) k + \frac {\epsilon k}2 = (0.75 + \epsilon) k$.
The stated running time is from \Cref{thm01} by the choice of $b \in O(1/\epsilon)$.
\end{proof}

\subsection{$k = 2$}
\label{subsec:general-bs2f}
In this subsection we design a direct algorithm for the {\sc BS$2$F} problem on general connected graphs.
Again let $G = (V, E, w)$ denote the edge non-negatively weighted graph.
The design ideas in the algorithm are the same, except that recursion is not needed,
because for $k = 2$, for each small edge set of the minimum spanning tree and the resulting bipartition by \Cref{op01},
only a minimum spanning tree is needed for each of the induced subgraph.
Nevertheless, to achieve the desired approximation ratio, the values of several parameters used in \Cref{alg03} need to be revised, as well as several lemmas and their proofs.
We first present the description of the algorithm in \Cref{alg04}, in which the {\em bound} $b = 20$ is on the size of the edge sets of the minimum spanning tree to be enumerated.

\begin{algorithm}[htb]
\caption{General-$2$-Forest $(G = (V, E, w), b = 20)$\label{alg04}}
\begin{algorithmic}[1]
\Require
	A connected graph $G = (V, E, w)$, an integer $b = 20$;
\Ensure
	A spanning $2$-forest of $G$.
\State Compute a minimum spanning tree $T$ of $G$;
\State convert $T$ into a spanning $2$-forest $\mcT$ by removing an edge, and record $\mcT$;\Comment{A candidate is guaranteed;}
\If {($\max_{T \in \mcT} w(T) = 0$)}
	\State\Return $\mcT$;
\EndIf
\ForAll {($D \subseteq E(T)$ with $1 \le |D| \le \min\{b, n-1\}$)}\Comment{Enumerate all small cuts;}
	\State $\{U_1, U_2\} = {\rm Bipartition}(T, D)$ by \Cref{op01};
	\If {(both $G[U_1]$ and $G[U_2]$ are connected)}
		\State record the $2$-forest $\{\MST(G[U_1]), \MST(G[U_2])\}$;
	\EndIf
\EndFor
\If {($T$ contains no $0.75$-balanced edge)}\Comment{Set $\beta = 0.75$;}
	\State orient the edges of $T$ as in \Cref{lemma07}, choose a sink $v$, and construct $H_v$ by Operation~\ref{op02};
	\ForAll {($L \in \{w(T)/6, w(T)/7\}$)}
		\If {($(H_v, p, c, L)$ is feasible)}
			\State call the $2$-approximation of \Cref{lemma06} with quota $L$ to obtain a tree $R$;
			\State prune $R$ by \Cref{lemma16} to satisfy $L \le p(R) < 2 L$;
			\State expand $R$ by \Cref{lemma09} to obtain and record a $2$-forest $\{T_0, T_1\}$;
		\EndIf
	\EndFor
\EndIf
\State\Return the recorded $2$-forest of minimum objective value.
\end{algorithmic}
\end{algorithm}

The next lemma is analogous to \Cref{lemma11}, but here to control the profit of a single auxiliary tree in an auxiliary graph without increasing its cost.
The pruning process is called inside the algorithm (Line 14 of \Cref{alg04}).

\begin{lemma}
\label{lemma16}
Given an auxiliary tree $R$ with $p(R) \ge L > 0$ in an auxiliary graph and every auxiliary vertex $C$ of $R$ satisfies $p(C) < 2 L$,
repeatedly delete a leaf vertex $C$ from $R$ whenever $p(R - C) \ge L$.
When no such deletion is possible, the resulting tree satisfies $L \le p(R) < 2 L$.
The pruning does not increase the cost of $R$.
\end{lemma}
\begin{proof}
One sees that every deletion preserves $R$ being a tree, preserves $p(R) \ge L$, decreases the number of vertices by one, and decreases the number of edges by one.
If there is only one vertex remained in $R$, then $L \le p(R) < 2 L$.
Otherwise, $R$ contains at least two leaves, and $p(R) \ge 2 L$ implies the lighter leaf can be deleted.
This proves the lemma.
\end{proof}

We first prove the correctness of \Cref{alg04} that it returns a feasible spanning $2$-forest in polynomial time in the next theorem,
and then analyze its performance.

\begin{theorem}
\label{thm03}
Given a connected graph $G = (V, E, w)$, \Cref{alg04} returns a spanning $2$-forest of $G$ in $O(n^b m \log n)$ time, where $b = 20$.
\end{theorem}
\begin{proof}
A minimum spanning tree $T$ can be computed in $O(m \log n)$ time, and if it contains only one or no edge of positive weight then the returned $2$-forest $\mcT$ (Line 4) is optimal.

Below we assume $w(T) > 0$.

In the for-loop (Lines 5--8) that enumerates all small edge sets $D \subseteq E(T)$,
a spanning $2$-forest is recorded in which the two trees are the minimum spanning trees of $G[U_1]$ and $G[U_2]$, respectively,
which are connected and $\{U_1, U_2\} =$ Bipartition$(T, D)$ by \Cref{op01}, and thus feasible.
The running time for this loop is $O(n^b m \log n))$, where $O(m)$ time is spent on generating the bipartition and the induced subgraphs,
and $O(m \log n)$ time for the subsequent minimum spanning trees.
Note that this for-loop records at least $n-1$ feasible solutions $T - e$ for each edge $e \in E(T)$.

At the non-existence of any $0.75$-balanced edge, by \Cref{lemma07} with $\beta = 0.75$ and \Cref{op02},
a sink $v$ can be determined and the auxiliary graph $H_v$ constructed in $O(m)$ time, and $p(C) < \frac 14 w(T)$ for every auxiliary vertex $C$ of $H_v$.
For either value $L \in \{\frac 16 w(T), \frac 17 w(T)\}$, we have $p(C) < \frac 14 w(T) < 2 L$.
Therefore, if the generated instance $(H_v, p, c, L)$ is feasible, then a tree $R$ with $p(R) \ge L$ is returned and subsequently pruned by \Cref{lemma16} so that $L \le p(R) < 2 L$.
The achieved $R$ is then expanded by \Cref{lemma09} to a spanning $2$-forest.
We note that the generated instance $(H_v, p, c, L)$ could be infeasible.
This process (Lines 9--15) takes time in a low order of polynomial in $m$ and $n$.

Since the for-loop dominates the running time, \Cref{alg04} overall runs in $O(n^{20} m \log n)$ time.
\end{proof}

For the performance analysis, let $\mcT^* = \{T^*_1, T^*_2\}$ denote a fixed optimal spanning $2$-forest of $G$ for discussion, and likewise $V^*_i = V(T^*_i)$ for $i = 1, 2$.
By \Cref{lemma02}, one sees if the size of the edge boundary of $V^*_1$ in the minimum spanning tree $T$, i.e., $r: = |\delta_T(V_1^*)|$, is at most $b$,
then \Cref{alg04} returns an optimal $2$-forest;
otherwise, the multigraph $G^*$ constructed right above \Cref{lemma04} using $T$ and $\mcT^*$ contains only $k = 2$ vertices with $r > b$ parallel edges,
and hence contains $r$ pairwise edge-disjoint spanning trees, and consequently \Cref{lemma04} states that $w(T) \le \frac {2r}{r-1} \OPT \le \frac {21}{10} \OPT$ using $b = 20$.

The next lemma goes a bit further than \Cref{lemma16} to possibly reduce the profit of a single auxiliary tree in an auxiliary graph while at the same time reduce the cost by half.

\begin{lemma}
\label{lemma17}
Given an auxiliary tree $R$ with $p(R) \ge 3 L > 0$ in an auxiliary graph, $R$ contains a subtree $S$ with $p(S) \ge L$ and $c(S) \le \frac 12 c(R)$.
\end{lemma}
\begin{proof}
If some vertex $C$ of $R$ has profit $p(C) \ge L$, then the single-vertex tree $S = \{C\}$ proves the lemma.

Assume henceforth every vertex of $R$ has profit less than $L$.
Root $R$ arbitrarily and choose a vertex $C$ farthest from the root such that the rooted subtree at $C$ has profit at least $L$.
It follows that $C$ has at least one child and the subtree rooted at every child of $C$ has profit less than $L$.
Starting with $C$, add subtrees rooted at its children one by one and stop when the accumulated profit reaches $L$ or greater, and denote the achieved subtree by $S_1$.
One sees that $L \le p(S_1) < 2 L$.
Let $S_2$ denote the subtree obtained from $R$ by deleting $V(S_1) \setminus \{C\}$.
The profit of $S_2$ is $p(S_2) = p(R) - p(S_1) + p(C) > 3 L - 2 L + p(C) \ge L$.

Note that $S_1$ and $S_2$ share exactly one common vertex $C$, and the edge sets $E(S_1)$ and $E(S_2)$ partition $E(R)$.
Therefore, $c(S_1) + c(S_2) = c(R)$.
Choosing the one between $S_1$ and $S_2$ with the smaller cost as $S$ proves the lemma.
\end{proof}

\begin{theorem}
\label{thm04}
\Cref{alg04} is a $7/4$-approximation algorithm for the {\sc BS$2$F} problem.
\end{theorem}
\begin{proof}
We have shown in \Cref{thm03} that \Cref{alg04} runs in polynomial time and returns a feasible spanning $2$-forest.
Let $T$ be the minimum spanning tree computed by the algorithm for the input edge nonnegatively weighted graph $G = (V, E, w)$.
From the earlier discussion, we assume $T$ contains at least two edges of positive weight.
We fix an optimal spanning $2$-forest $\{T_1^*, T_2^*\}$ for discussion, and assume $r = |\delta_T(V_1^*)| \ge b + 1 = 21$, which implies $w(T) \le \frac {21}{10} \OPT$.
We distinguish the two cases on whether a $0.75$-balanced edge of $T$ exists or not.

{\bf Case 1.} There is a $0.75$-balanced edge $e$ in $T$.
In this case, since $D = \{e\}$ is enumerated in the for-loop, at which the spanning $2$-forest $\mcT = \{T_1, T_2\}$ has its objective value at most
$0.75 w(T) \le \frac {63}{40} \OPT \le \frac 74 \OPT$.

{\bf Case 2.} There is no $0.75$-balanced edge in $T$.
In this case, we have proved in \Cref{thm03} that a sink $v$ can be determined and the auxiliary graph $H_v$ constructed in which $p(C) < \frac 14 w(T)$ for every auxiliary vertex $C$.
Without loss of generality, we assume $v \in V_1^*$.
By \Cref{lemma05}, using $k = 2$ and $i_0 = 1$, $H_v$ contains a tree $Q_2$ with $c(Q_2) \le \OPT$ and $p(Q_2) \ge w(T) - \OPT$.

{\bf Subcase 2.1.} $\OPT \le \frac 12 w(T)$.
Consider when $L = w(T)/6$ is used as the quota.
Since $p(Q_2) \ge w(T) - \OPT \ge \frac 12 w(T) = 3 L$, $(H_v, p, c, L)$ is a feasible {\sc QST} instance and \Cref{lemma17} ensures a feasible tree of cost at most $\frac 12 \OPT$.
Therefore, \Cref{lemma06} returns a tree $R$ with $p(R) \ge L$ and $c(R) \le 2 \cdot \frac 12 \OPT = \OPT$.
Subsequently, \Cref{lemma16} prunes $R$ to satisfy $\frac 16 w(T) \le p(R) < \frac 13 w(T)$ and $c(R) \le \OPT$.
Lastly, \Cref{lemma09} expands $R$ to obtain a tree $T_1$ in $G$ with $w(T_1) \le p(R) + c(R) < \frac {17}{10} \OPT \le \frac 74 \OPT$ and
the residual tree $T_0$ with $w(T_0) = w(T) - p(R) \le \frac 56 w(T) \le \frac 74 \OPT$.

{\bf Subcase 2.2.} $\frac 12 w(T) < \OPT \le \frac 47 w(T)$.
Consider when $L = w(T)/7$ is used as the quota.
Since $p(Q_2) \ge w(T) - \OPT \ge \frac 37 w(T) = 3 L$, the same as in Subcase~2.1, \Cref{lemma06} returns a tree $R$ with $p(R) \ge L$ and $c(R) \le \OPT$.
This time, \Cref{lemma16} prunes $R$ to satisfy $\frac 17 w(T) \le p(R) < \frac 27 w(T)$ and $c(R) \le \OPT$.
Lastly, \Cref{lemma09} expands $R$ to $T_1$ with $w(T_1) \le p(R) + c(R) < \frac {11}7 \OPT < \frac 74 \OPT$ and
the residual tree $T_0$ with $w(T_0) \le \frac 67 w(T) < \frac {12}7 \OPT < \frac 74 \OPT$.

{\bf Subcase 2.3.} $\OPT > \frac 47 w(T)$.
Since $w(T) < \frac 74 \OPT$, $T - e$ for any edge $e$ of $T$ gives rise to a spanning $2$-forest in which both trees have weight at most $w(T) < \frac 74 \OPT$.

We thus have proved that in every case, Algorithm~\ref{alg04} records a spanning $2$-forest of objective value at most $\frac 74 \OPT$.
This proves the approximation ratio.
\end{proof}

\section{Approximation algorithms for metric complete graphs}
\label{sec:metric-complete}
This section is devoted to the special case where the input graph is complete and the edge weights satisfy triangle inequality,
which models many real-life applications.
We fix such a graph $G = (V, w)$ for consideration, where the nonnegative edge weight function satisfies $w(uv) \le w(uw) + w(wv)$ for any three vertices $u, v, w \in V$.
Additionally, if needed, we multiply the function by a positive integer so that all edge weights, as well as $\frac 12 \OPT$, become integers,
that is, we hereafter assume all edge weights are nonnegative integers.

Recall the closely related min-max tree cover problem to find a collection of $k$ trees in the given edge nonnegatively weighted graph such that
every vertex appears in at least one tree and the maximum weight of the trees is minimized.
We note that in this problem, the input graph is the same as what we have considered in the last section, that is,
the graph is not necessarily complete, neither the edge weights satisfy triangle inequality.
Khani and Salavatipour~\cite{KS14} propose a $3$-approximation algorithm, improving two different $4$-approximation algorithms by Arkin et al.~\cite{AHL06} and Even et al.~\cite{EGK04}.
We remind the readers the key difference between our problem and the min-max tree cover problem,
that we require the returned $k$ tree vertex-disjoint while they allow trees to overlap.

We explain first a simple $4$-approximation algorithm for our problem based on the {\em path construction} of Arkin et al.~\cite{AHL06},
and then adapt the {\em component-and-matching} framework of Khani and Salavatipour~\cite{KS14} to design a $3$-approximation algorithm.

\begin{definition}
\label{def02}
Given a $\lambda \ge 0$, $E_\lambda$ denotes the subset of edges of $G$ of weight at most $\lambda$, and $G_\lambda = (V, E_\lambda)$.
\end{definition}

Trivially, if the graph $G_0 = (V, E_0)$, i.e., $\lambda = 0$, contains $k$ connected components or less,
then $\OPT = 0$ and an optimal spanning $k$-forest can be constructed by taking a spanning tree for each component followed by removing some more edges to achieve $k$ trees.
We therefore assume in the rest of the section that $\OPT > 0$.

\subsection{A simple $4$-approximation}
Fix a $\lambda > 0$ and consider a connected component $C$ of the spanning subgraph $G_\lambda$.
If $\OPT \le \lambda$, then every optimal tree lies in a connected component of $G_\lambda$.
Assume $r$ optimal trees lie in $C$, then they can be joined by $r-1$ edges of $C$ to form a spanning tree of $C$, and thus $w(\MST(G[C])) \le (2r - 1) \lambda$.
Doubling this MST and shortcutting give a Hamiltonian path $P$ in $G[C]$ with $w(P) < 4r \lambda$.

Starting from one end of $P$, the shortest prefix with weight exceeding $4\lambda$ is cut off from $P$,
and then its last edge is discarded from the prefix to give a path of weight at most $4\lambda$;
and the process repeats until the weight of the remaining $P$ becomes no greater than $4\lambda$.
This way, the Hamiltonian path $P$ in $G[C]$ gives rise to at most $\max\{\lceil w(P)/(4\lambda) \rceil, 1\} \le r$ vertex-disjoint paths, each of weight at most $4\lambda$.
Unionizing these paths over all the components of $G_\lambda$ gives at most $k$ paths, which lead to a spanning collection of exactly $k$ vertex-disjoint paths.

Since $\OPT \in (0, w(\MST(G)))$, we use binary search for $\lambda$;
the final successful value of $\lambda$, i.e., by returning $k$ paths or less, is at most $\OPT$, and thus the associated solution has objective value at most $4 \cdot \OPT$.
This is a polynomial-time $4$-approximation algorithm for {\sc BS$k$F} on metric complete graphs.

\subsection{A $3$-approximation}
We fix a $\lambda > 0$ for discussion, of which the final value will be determined later also by binary search.

\begin{definition}[Light and heavy components]
\label{def03}
Let $C_1, \dots, C_{\ell + h}$ denote the connected components of $G_\lambda$.
For every component $C_i$, fix a minimum spanning tree of $G[C_i]$.
A component is {\em light} if $w(\MST(G[C_i])) \le 2\lambda$, or {\em heavy} otherwise.
The light components are indexed as $C_1, \dots, C_\ell$ and the heavy components as $C_{\ell+1}, \dots, C_{\ell+h}$.

For a light component $C_i$, if there is an edge of weight at most $2\lambda$ connecting $C_i$ to a heavy component,
then define $A(C_i)$ as $w(\MST(G[C_i]))$ plus the minimum weight of such an edge;
a heavy component attaining this minimum weight edge is called a {\em nearest} heavy component of $C_i$.
If no such edge exists, set $A(C_i) = +\infty$.
\end{definition}

\begin{definition}[Matching graph]
\label{def04}
Fix two integers $n_0 \ge 0$ and $n_h \ge 0$.
The {\em matching graph} $G_M$ associated with $G_\lambda$ contains a distinct {\em regular node} for every light component $C_i$ of $G_\lambda$,
$n_0$ {\em null nodes}, and $n_h$ {\em heavy nodes}.
Two regular nodes are adjacent if the minimum edge weight between the two corresponding light components is at most $2\lambda$,
and the edge between them has cost $0$ with which a minimum-weight original edge between the two components is stored.
Every null node is adjacent to every regular node by an edge of cost $0$.
If $A(C_i) < +\infty$, the regular node corresponding to $C_i$ is adjacent to every heavy node by an edge of cost $A(C_i)$,
with which a nearest heavy component in $G_\lambda$ and a minimum-weight original edge between the two components are stored.
\end{definition}

\begin{lemma}
\label{lemma18}
Given a tree $R$ in $G_\lambda$, one can compute in polynomial time a partition of $V(R)$ into nonempty sets $S_1, \dots, S_q$,
together with a minimum spanning tree of each induced subgraph $G[S_i]$,
such that $q \le \max\{\lfloor w(R)/\lambda \rfloor, 1\}$ and $w(\MST(G[S_i])) < 2 \lambda$ for every $i \in \{1, \dots, q\}$.
\end{lemma}
\begin{proof}
If $w(R) < 2 \lambda$, then the lemma is proven automatically.

When $w(R) \ge 2 \lambda$, root $R$ arbitrarily and find a farthest vertex $v$ from the root so that the subtree rooted at $v$ has weight at least $\lambda$.
Let the child branches be $B_1, \dots, B_d$, where $d \ge 1$ and each $B_i$ contains the edge $vu_i$ to the corresponding child $u_i$ and the subtree rooted at $u_i$.
Order these branches so that $w(B_1) \ge \dots \ge w(B_d)$.
We distinguish two cases.

{\bf Case 1.} $w(B_1) \ge \lambda$.
Noting that $w(B_1) - w(vu_1) < \lambda$ by the definition of $v$, $w(B_1) < 2 \lambda$.
Let $V(B_1) \setminus \{v\}$ be a set $S$,
then $w(\MST(G[S])) < \lambda$ and then repeat the process on the remainder tree $R[V(R) \setminus S]$ of which the weight is reduced by $w(B_1) \ge \lambda$.

{\bf Case 2.} $w(B_1) < \lambda$.
Let $r$ be the smallest index for which $A = \sum_{i=1}^r w(B_i) \ge \lambda$.
Then $r \ge 2$ and $A - w(B_r) < \lambda$.
Let $\cup_{i=1}^r V(B_i) \setminus \{v\}$ be a set $S$.
By triangle inequality, the minimum spanning tree on the children $u_1, \dots, u_r$ has weight no more $\sum_{i=1}^r w(vu_i) + (r-2) \min_{i=1}^r w(vu_i)$,
and consequently the minimum spanning tree on $S$ has weight at most $A + (r-2) \min_{i=1}^r w(vu_i) \le A + (r-2) w(B_r) \le 2 (A - w(B_r)) < 2 \lambda$.
One then repeat the process on the remainder tree $R[V(R) \setminus S]$ of which the weight is reduced by $A \ge \lambda$.

In summary, whenever a set $S$ is constructed,
its minimum spanning tree in $G[S]$ has weight less than $2 \lambda$ and the weight of the remainder tree $R$ is reduced by at least $\lambda$.
The time spent is for tree traversal and minimum spanning tree construction, which is in $O(m \log n)$.
Since the last set construction is done when $w(R) \ge 2 \lambda$, the number of constructed sets plus the last remaining $V(R)$ is at most $w(R)/\lambda$.
This proves the lemma.
\end{proof}

Using \Cref{lemma18}, for each heavy component $C_i$ of $G_\lambda$,
its minimum spanning tree has $w(\MST(G[C_i])) > 2 \lambda$ but can be partitioned into trees of weight less than $2 \lambda$
with the help of some possible edges of weight greater than $\lambda$.
Nevertheless, this could produce too many trees and thus we will need to consider light components and heavy components jointly;
at the same time, we will need to increase the upper bound of $2 \lambda$.
The following algorithm in \Cref{alg05} does this by identifying light components that can be paired up, that can be attached to their nearest heavy component, and that can be left alone,
through enumerating $(n_0, n_h)$, where $n_0$ is the number of light components left alone and $n_h$ is the number of light components attached to their nearest heavy component,
to construct the associated matching graph $G_M$ followed by computing a minimum-cost perfect matching $M$ of $G_M$.
If successful, the algorithm returns a spanning $k$-forest in which each tree has weight no greater than $6 \lambda$.

\begin{algorithm}[htb]
\caption{Metric-$k$-Forest $(G = (V, w), k, \lambda)$\label{alg05}}
\begin{algorithmic}[1]
\Require
	A metric complete graph $G = (V, w)$, an integer $k \ge 2$, and a parameter $\lambda > 0$;
\Ensure
	A spanning $k$-forest of $G$, or {\em failure} if no such forest is found.
\State Construct $G_\lambda$ and its light components $C_1, \dots, C_\ell$ and heavy components $C_{\ell+1}, \dots,C_{\ell+h}$ (\Cref{def03});
\ForAll{($(n_0, n_h)$ with $0 \le n_0, 0 \le n_h, n_0 + n_h \le \ell$, and $\ell - n_0 - n_h$ even)}
	\State construct matching graph $G_M$ (\Cref{def04});
	\State compute a minimum-cost perfect matching $M$ of $G_M$, if exists;
	\If {($M$ exists)}
		\State initialize an empty collection $S$ of trees;
		\ForAll{(regular node $C_i$ matched to a null node)}
			\State add $\MST(G[C_i])$ to $S$;
		\EndFor
		\ForAll{(two matched regular nodes $C_i$ and $C_j$)}
			\State join $\MST(G[C_i])$ and $\MST(G[C_j])$ by the stored minimum-weight edge;
			\State add the resulting tree to $S$;
		\EndFor
		\ForAll{(regular node $C_i$ matched to a heavy node)}
			\State attach $\MST(G[C_i])$ to its stored nearest heavy component by the stored edge;
		\EndFor
		\For{($j = \ell+1$ to $\ell+h$)}
			\State set $T_j$ to be the tree formed by $\MST(G[C_j])$ and all light-component MSTs attached to $C_j$;
			\State apply \Cref{lemma18} to $T_j$ in $G_{3 \lambda}$, add the returned MSTs to $S$;
		\EndFor
		\If{($|S| \le k$)}\Comment{The algorithm succeeds and returns a spanning $k$-forest;}
			\State repeatedly delete edges from trees of $S$ until $|S| = k$;
			\State \Return $S$;
		\EndIf
	\EndIf
\EndFor
\State \Return {\em failure}.
\end{algorithmic}
\end{algorithm}

\begin{lemma}
\label{lemma19}
Given a metric complete graph $G = (V, w)$ and a parameter $\lambda > 0$, Algorithm~\ref{alg05} runs in $O(n^5)$ time,
and if Algorithm~\ref{alg05} succeeds, then it returns a spanning $k$-forest of $G$ in which every tree has weight at most $6 \lambda$.
\end{lemma}
\begin{proof}
Note that if Algorithm~\ref{alg05} succeeds, then for some pair $(n_0, n_h)$ the constructed matching graph $G_M$ admits a perfect matching $M$,
in which a regular node is matched to a null node, to another regular node, or to a heavy node,
meaning that the corresponding light component is left alone, is paired with another light component, or is attached to its nearest heavy component.

Consider the weights of the trees before the final stage of edge deletions to create exactly $k$ trees (Lines 17--18), in the three separate cases.

{\bf Case 1.} A regular node $C_i$ is matched to a null node.
In this case, the constructed tree is $\MST(G[C_i])$ which has weight at most $2 \lambda$ by definition.

{\bf Case 2.} Two regular nodes $C_i$ and $C_j$ are matched.
Both component MSTs $\MST(G[C_i])$ and $\MST(G[C_j])$ have weight at most $2 \lambda$ and the stored connecting edge has weight at most $2 \lambda$ too.
Therefore, the resulting tree has weight at most $6 \lambda$. 

{\bf Case 3.} A regular node $C_i$ is matched to a heavy node.
Consider the intermediate tree $T_j$ for which $C_j$ is the nearest heavy component of $C_i$.
That is, $T_j$ is $\MST(G[C_j])$ attached with $\MST(G[C_i])$ for those regular nodes $C_i$ matched to a heavy node and $C_j$ is its nearest heavy component.
The edges inside $\MST(G[C_j])$ and $\MST(G[C_i])$ have weight at most $\lambda$,
and every light-to-heavy attachment edge has weight at most $2 \lambda$.
Therefore, $T_j$ is in $G_{2 \lambda}$ and consequently in $G_{3 \lambda}$ as well.
\Cref{lemma18} applies and partitions $V(T_j)$ to give $q$ trees each of weight less than $6 \lambda$.

We thus prove that the constructed trees are vertex-disjoint, together they contain all the vertices of $V$, and each tree has weight at most $6 \lambda$.
As Algorithm~\ref{alg05} succeeds for this pair $(n_0, n_h)$,
the number of trees is at most $k$, and the returned a spanning $k$-forest is done by possibly deleting more tree edges to make the number exactly $k$.

In terms of the running time, one sees that the construction of $G_\lambda$ and light-vs-heavy component determination in Line 1 take $O(n^2)$ time.
The for-loop enumerates $(n_0, n_h)$ pairs and there are $O(n^2)$ of them;
in each iteration, the most time consuming step is the minimum-cost perfect matching computation, which can be done in $O(n^3)$ time.
Therefore the overall running time is within $O(n^5)$, where $n = |V|$.
\end{proof}

With \Cref{alg05}, we apply a similar binary search for $\lambda$ in the range $(0, w(\MST(G)))$,
and the last successful call to \Cref{alg05} gives the final spanning $k$-forest for $G$.
A description of the algorithm is in \Cref{alg06}.

\begin{algorithm}[htb]
\caption{Metric-BS$k$F $(G = (V, w), k)$\label{alg06}}
\begin{algorithmic}[1]
\Require
	A metric complete graph $G = (V, w)$, an integer $k \ge 2$;
\Ensure
	A spanning $k$-forest of $G$.
\State Compute a minimum spanning tree $T$ of $G$, store the $k$-forest for $w(T)$ by removing $k-1$ edges from $T$;
\State set $LB \gets 0$, $UB \gets w(T)$;
\While{($UB > LB + 1$)}
	\State set $\lambda \gets \lfloor (LB + UB)/2 \rfloor$;
	\State call \Cref{alg05} on the instance $(G, k, \lambda)$;
	\If{(successful)}
		\State set $UB \gets \lambda$, store the returned $k$-forest for $UB$;
	\Else
		\State set $LB \gets \lambda$;
	\EndIf
\EndWhile
\State \Return the $k$-forest stored for $UB$.
\end{algorithmic}
\end{algorithm}

\begin{theorem}
\label{thm05}
Given a metric complete graph $G = (V, w)$, where $w: V^2 \to \mathbb{Z}_{\ge 0}$, \Cref{alg06} returns a feasible spanning $k$-forest in $O(n^5 \log (w(\MST(G)) + 1))$.
\end{theorem}
\begin{proof}
Note that the spanning $k$-forest stored for $w(T)$ by removing $k-1$ edges from the minimum spanning tree $T$ in Line 1 is feasible,
which guarantees the $k$-forest stored for $UB$ if $UB$ is not updated in the binary search by the while-loop.
Every time $UB$ is updated, calling \Cref{alg05} on the instance $(G, k, UB)$ succeeds and returns a spanning $k$-forest by \Cref{lemma19}.
Since the lower bound $LB$ is set at $0$, the total number of iterations in the while-loop is at most $\log (w(\MST(G)) + 1)$,
and thus by \Cref{lemma19} again the overall running time is in $O(n^5 \log (w(\MST(G)) + 1))$.
\end{proof}

By \Cref{thm05}, \Cref{alg06} stops at a specific $\lambda > 0$ and returns a spanning $k$-forest $\mcT$ stored for $\lambda$, while calling \Cref{alg05} using $\lambda - 1$ fails.
For the performance analysis, we stick with this specific $\lambda$.
We fix an optimal spanning $k$-forest $\mcT^* = \{T_1^*, \dots, T_k^*\}$ with objective value $\OPT$ for discussion,
and the trees inside $\mcT^*$ are called optimal trees.
Since $\mcT$ has objective value at most $6 \lambda$ by \Cref{lemma19}, to prove the approximation ratio of $3$, we focus on the case where $\OPT \le 2 \lambda$.

\begin{definition}[Optimal-tree types]
\label{def05}
Assume $\OPT \le 2 \lambda$, then every optimal tree contains at most one edge of weight greater than $\lambda$ and therefore intersects at most two components of $G_\lambda$.
\begin{itemize}
\parskip=0pt
\item
	An optimal tree that intersects two components is called a {\em broken} tree;
	the unique edge connecting the two components is its {\em bridge} edge, and deleting the bridge edge gives two {\em broken subtrees}.
\item
	An optimal tree is called a {\em light} tree if every component it intersects is light,
	a {\em heavy} tree if every component it intersects is heavy,
	and a {\em mixed} tree if one broken subtree is in a light component and the other in a heavy component.
	Let $k_\ell$, $k_h$, and $k_m$ denote the numbers of light, heavy, and mixed optimal trees, respectively;
	so $k_\ell + k_h + k_m = k$.
\end{itemize}
\end{definition}

\begin{lemma}
\label{lemma20}
Assume $\OPT \le 2 \lambda$.
Then the for-loop of \Cref{alg05} enumerates a pair $(n_0, n_h)$ for which the matching graph $G_M$ has a perfect matching.
For the minimum-cost perfect matching $M$ computed for $G_M$ for this pair, the light components not attached to any heavy components produce at most $k_\ell$ trees.
Moreover, the cost of this matching $M$ is at most the sum of the weight of the light broken subtree plus the weight of its bridge edge in all the mixed optimal trees.
\end{lemma}
\begin{proof}
We prove the lemma by construction.

Construct a multigraph $G_\ell^*$ in which a node corresponds to a light component of $G_\lambda$.
For every light optimal tree $T_i^*$, add an edge between the two corresponding nodes if $T_i^*$ is broken, or otherwise add a loop at the corresponding node.
This way, $G_\ell^*$ has exactly $\ell$ nodes and exactly $k_\ell$ edges including loops.

A node of $G_\ell^*$ is {\em isolated} if there is no edge incident to it, and let $I$ be the set of the isolated nodes in $G_\ell^*$.
Let $M^*$ be a maximum matching of $G_\ell^*$, and let $U$ be the set of non-isolated nodes not saturated by $M^*$.

For each node of $U$, use its loop if one exists;
otherwise choose an incident edge, of which the other end-node is saturated by $M^*$.
These chosen edges are distinct for the nodes of $U$ and none of them is in $M^*$.
Therefore, $|M^*| + |U| \le k_\ell$.

For each node of $I$ and the corresponding light component $C_i$ of $G_\lambda$, every vertex of $C_i$ belongs to the light broken subtree of a mixed tree.
Suppose there are $r$ such broken subtrees in $C_i$, then they can be interconnected inside $C_i$ by $r-1$ edges into a tree.
Let the corresponding bridge-edge weights be $b_1, \dots, b_r$, and $b_{\min}$ be their minimum.
Since $\lambda < b_j \le \OPT \le 2\lambda$ for every $j \in [r]$, by \Cref{def03},
$A(C_i)$ is finite and is at most the sum of the light broken-subtree weights, $(r-1) \lambda$, and $b_{\min}$,
which is less than or equal to the sum of the light broken-subtree weights plus $\sum_{j=1}^r b_j$.
Therefore, the sum of $A(C_i)$ over all isolated light components $C_i$ is upper bounded by the total,
over mixed optimal trees, of the light broken-subtree weight plus the bridge-edge weight.

Now we set $n_0 = |U|$ and $n_h = |I|$, and subsequently the matching graph $G_M$.
We construct a perfect matching in $G_M$ as follows:
Match the regular nodes corresponding to $U$ to the $n_0$ null nodes,
match the regular nodes corresponding to $I$ to the $n_h$ heavy nodes,
and pair the remaining regular nodes according to the matching $M^*$ of $G_\ell^*$,
for which the regular-regular edges are guaranteed by the edges of $M^*$, and the regular-heavy edges are guaranteed by the mixed trees.
Furthermore, $G_M$ has a total cost equal to the sum of $A(C_i)$ over all isolated light components $C_i$, which is upper bounded as the stated.

\Cref{alg05} computes a minimum-cost perfect matching for $G_M$ for this pair $(n_0, n_h)$, which has a lower or the same cost.
Furthermore, the identity $\ell = 2 |M^*| + n_0 + n_h$ shows that every perfect matching of $G_M$ has exactly $|M^*|$ regular-regular edges.
That is, the light components not attached to their nearest heavy component produce exactly $n_0 + |M^*| \le k_\ell$ trees.
This proves the lemma.
\end{proof}

\begin{lemma}
\label{lemma21}
Assume $\OPT \le 2\lambda$.
For the parameter pair $(n_0, n_h)$ in \Cref{lemma20},
let $T_{\ell+1}, \dots, T_{\ell+h}$ be the trees formed by $\MST(G[C_j])$ and all the light-component MSTs attached to the heavy component $C_j$ in \Cref{alg05}.
Then $\sum_{j = \ell+1}^{\ell+h} w(T_j) \le 3 (k_h + k_m) \lambda - h \lambda$.
After applying \Cref{lemma18} to each of them in $G_{3\lambda}$, these trees $T_{\ell+1}, \dots, T_{\ell+h}$ produce at most $k_h + k_m$ trees each of weight less than $6 \lambda$.
\end{lemma}
\begin{proof}
Recall \Cref{def05} on the types of optimal trees.

Let $p$ be the number of broken heavy optimal trees and $q$ be the number of non-broken heavy optimal trees (i.e., contained in a single heavy component);
we have $p + q = k_h$.
Since a broken heavy tree has a bridge edge of weight greater than $\lambda$, the total weight of its two broken subtrees is less than $\lambda$.
It follows that the total weight of heavy optimal trees remained inside heavy components of $G_\lambda$ is at most $(p + 2q) \lambda$.

A mixed tree is partitioned into three parts, the heavy broken subtree, the light broken subtree, and the bridge edge.
The weight of the light broken subtree and the weight of the bridge edge are counted towards the upper bound on $A(C_i)$ for the light component $C_i$ intersected by the mix tree.
Thus all the mixed trees contribute at most $2k_m \lambda$, of which the total weight of the heavy broken subtrees remained inside heavy components of $G_\lambda$
while the other part counted towards the upper bound on the cost of the perfect matching of $G_M$ in \Cref{lemma20}.

Inside all the $h$ heavy components of $G_\lambda$ there are $2p + q + k_m$ optimal trees or broken subtrees,
which can be interconnected using at most $2p + q + k_m - h$ additional edges into $h$ spanning trees each for a heavy component.
Hence the total weight of the heavy-component MSTs plus the sum of $A(C_i)$ across the light components $C_i$ intersected by a mix tree is
\[
\sum_{j = \ell+1}^{\ell+h} w(T_j) \le (p + 2q) \lambda + 2k_m \lambda + (2p + q + k_m - h) \lambda = 3(k_h + k_m) \lambda - h\lambda.
\]
This proves the first half of the lemma.

Let $q_j$ be the number of trees produced by \Cref{lemma18} on $T_j$ in $G_{3\lambda}$.
If $q_j \ge 2$, then the lemma gives $w(T_j) \ge q_j \cdot 3\lambda = 3q_j \lambda$.
If $q_j = 1$, the intermediate tree contains the MST of a heavy component and therefore has weight greater than $2 \lambda = 3q_j \lambda - \lambda$.
Therefore, in either case, $w(T_j) \ge 3q_j \lambda - \lambda$.
Summing over all the $h$ heavy components and using the preceding upper bound gives $3 \lambda \sum_{j=1}^h q_j - h \lambda \le 3(k_h + k_m) \lambda - h\lambda$,
and hence $\sum_{j=1}^h q_j \le k_h + k_m$.
\end{proof}

Combining the above two lemmas presents the success condition for a fixed value of $\lambda$.

\begin{theorem}
\label{thm06}
If $\OPT \le 2\lambda$, then \Cref{alg05} succeeds and returns a spanning $k$-forest of value at most $6\lambda$.
\end{theorem}
\begin{proof}
By \Cref{lemma20}, \Cref{alg05} enumerates a parameter pair for which a perfect matching exists and the light components not attached to any heavy component produce at most $k_\ell$ trees.
By \Cref{lemma21}, the heavy components and their attached light components produces at most $k_h + k_m$ trees.
That is, before the final step of possible edge deletions, the spanning forest contains at most $k_\ell + k_h + k_m = k$ trees.
\Cref{alg05} thus succeeds for this parameter pair with the weight of the trees bounded by $6 \lambda$ by \Cref{lemma19}.
\end{proof}

\begin{theorem}
\label{thm07}
\Cref{alg06} is a polynomial-time $3$-approximation algorithm for the {\sc BS$k$F} problem on metric complete graphs.
\end{theorem}
\begin{proof}
Given an input metric complete graph $G = (V, w)$, as we stated at the beginning of the section that whether or not $\OPT = 0$,
and if so then an optimal spanning $k$-forest can be returned, in polynomial time, we consider the remaining case where $\OPT > 0$.
We also assume that all the edge weights are integers.

For any $\lambda \ge \frac 12 \OPT$, i.e., $\OPT \le 2 \lambda$,
\Cref{thm06} states that \Cref{alg05} succeeds and returns a spanning $k$-forest of value at most $6\lambda$ for $\lambda$.
For the initial upper bound $UB = w(T)$, the spanning $k$-forest obtained by deleting $k-1$ edges of the minimum spanning tree (Line 1 of \Cref{alg06}) has its objective value at most $UB$
and is stored for $UB$.
Therefore, during the binary search, $UB$ is always a successful value, while $LB$ is either $0$ or a value at which \Cref{alg05} failed.
It follows that when the binary search, i.e., the while-loop, terminates,
$UB \le \frac 12 \OPT$ and the spanning $k$-forest stored for it has its objective value at most $6 \cdot UB \le 3 \cdot \OPT$.
This proves the approximation ratio of $3$.

The polynomial running time has been proven in \Cref{thm05}.
\end{proof}

\section{Hardness of approximation}
\label{sec:computational-complexity}
Chleb\'ik and Chleb\'ikov\'a~\cite{CC06} proved a number of approximation lower bounds for low degree graph problems, including the {\em minimum vertex cover} ({\sc VC}) problem.
They started with the following restricted version of {\em maximum linear equations} over $\mathbb{Z}_2$ with three variables per equation,
denoted by {\sc E$[k, k, k]$-Max-E$3$-Lin-$2$}, of which an instance is a system of linear equations over $\mathbb{Z}_2$ with exactly $3$ (distinct) variables in each equation,
and each variable occurs exactly $k$ times as the first variable, $k$ times as the second variable, and $k$ times as the third variable in equations;
the objective is to find an assignment to the variables to satisfy the maximum number of equations.

H{\aa}stad~\cite{Has01} showed that for every $\epsilon \in (0, 0.25)$, there is an integer $k(\epsilon)$ such that for all $k \ge k(\epsilon)$,
it is NP-hard to decide an instance of {\sc E$[k, k, k]$-Max-E$3$-Lin-$2$} on whether a fraction of more than $1 - \epsilon$ or less than $0.5 + \epsilon$ of all equations
is satisfied by an optimal assignment.

Given a graph $H = (U, F)$, a {\em vertex cover} is a set $S \subseteq U$ so that every edge of $F$ is incident to some vertex of $S$.
Using the so-called {\em consistency gadgets}, Chleb\'ik and Chleb\'ikov\'a~\cite{CC06} proved the following lemma on the {\sc VC} problem.

\begin{lemma}
\label{lemma22}
For every $\epsilon \in (0, 0.25)$, there exists a fixed nonnegative rational number $\mu \le 20.7$ for which the following problem is NP-hard:
Given a graph $H = (U, F)$ of maximum degree $5$ and an integer $q \ge 1$ satisfying $|U| = (2\mu + 6) q$ and $|F| \le (4\mu + 13) q$,
distinguish between the cases where $H$ has a vertex cover of size at most $(\mu + 4 + \epsilon) q$ and where every vertex cover has size at least $(\mu + 4.5 - \epsilon) q$.
\end{lemma}
\begin{proof}
Given an instance $I$ of {\sc E$[k, k, k]$-Max-E$3$-Lin-$2$} with $q = mk$ equations on $m$ variables,
In the proof of Chleb\'ik and Chleb\'ikov\'a~\cite[Theorem~17]{CC06} for the {\sc VC} problem on graphs of maximum degree $5$ ({\sc Max-$5$-VC} in \cite{CC06}),
a graph $H = (U, F)$ of maximum degree $5$ is constructed with $|U| = (2\mu + 6) q$,
where $\mu$ is a quantity defined by the parameters of the {\em consistency gadget} independent of $m$ and $k$, and $\mu \le 20.7$.
It is shown that the minimum size of the vertex cover in $H$ is $(\mu + 5 - \OPT(I)) q$, where $\OPT(I)$ denotes the maximum satisfiable fraction of equations in $I$.
Since by H{\aa}stad~\cite{Has01} it is NP-hard to decide whether $\OPT(I) \ge 1 - \epsilon$ or $\OPT(I) \le 0.5 + \epsilon$,
the lemma is proved except the number of edges in $H$.

Chleb\'ik and Chleb\'ikov\'a~\cite[Theorem~17]{CC06} also proved the hardness for {\sc Max-$4$-VC},
in which the same definition using the parameters of the consistency gadget gives $\mu + 1 \le 21.7$.
The constructed graph has a maximum degree $4$ and contains $(2 \mu + 8) q$ vertices, and thus has at most $2 (2\mu + 8) q$ edges.
The above degree-$5$ graph $H = (U, F)$ is obtained from this degree-$4$ graph
by replacing each $4$-cycle associated with a contact in the second group of the consistency gadget by a single edge,
and there are precisely $q$ such replacements since each of the $m$ variable occurs exactly once in the second position for a total of $k$ occurrences.
One sees that each replacement removes two vertices and replaces four internal edges by one,
and consequently $|U| = (2\mu + 6) q$ and $|F| \le (4\mu + 16 - 3) q = (4\mu + 13) q$.
\end{proof}

\subsection{Metric complete graphs}
We prove a lower bound on approximation ratios for the {\sc BS$k$F} problem on metric complete graphs.
In particular, when $k = 2$, all the edge weights can be restricted in $\{1, 2, 3\}$.
Our reduction starts from the {\sc VC} problem, of which an instance consists of a graph $H = (U, F)$, which is a fixed number of disjoint copies of the degree-$5$ graph in \Cref{lemma22}.

\begin{lemma}
\label{lemma23}
Let $H = (U, F)$ be a graph with $N = |U|$ and $M = |F|$, and $K$ be an integer satisfying $1 \le K \le N$.
Suppose that $2M + 3K - N - 2 \ge 0$ and $B = 2(M + K - 1) > 0$.
A metric complete graph $G = (V, w)$ with weights in $\{1, 2, 3\}$ satisfying the following properties can be constructed in polynomial time:
\begin{itemize}
\parskip=0pt
\item
	If $H$ has a vertex cover of size at most $K$, then $\OPT = B$ for $G$;
	in particular, $G$ has a spanning $2$-forest in which both trees have weight $B$.
\item
	Given any spanning $2$-forest of $G$ with objective value $C$,
	one can either conclude that $C \ge B + K$ or construct a vertex cover of $H$ of size at most $K + \frac 72 (C - B)$ in polynomial time.
\end{itemize}
\end{lemma}
\begin{proof}
We construct the graph as follows:
\begin{itemize}
\parskip=0pt
\item
	The vertex set $V$:
	For every $v \in U$ create a vertex $x_v$, $X = \{x_v: v \in U\}$, and for every $e \in F$ create a vertex $y_e$, $Y = \{y_e: e \in F\}$.
	Add a vertex $v_0$ and a set $Z$ of $2M + 3K - N - 2$ distinct vertices.
	Then, $V = \{v_0\} \cup X \cup Y \cup Z$.
\item
	The edge weights:
	Assign $w(v_0x) = 1$ for $x \in X \cup Z$;
	$w(xy) = 2$ whenever $x, y \in X \cup Z$ are distinct, or whenever $x = x_v \in X$, $y = y_e \in Y$, and $v$ is an endpoint of $e$;
	and $w(e) = 3$ for all the remaining edges.
\end{itemize}
Note that $|V| = 1 + N + M + |Z|$, and thus construction takes polynomial time.
Given that all edge weights are in $\{1, 2, 3\}$, we only need to verify non-existence of a triangle of edge weights $1$-$1$-$3$ for satisfying the triangle inequality.
This is simple since every weight-$1$ edge is incident to $v_0$, and their other endpoints are connected by weight-$2$ edges.

We next prove the two properties.

Suppose $H$ has a vertex cover $S$ of size at most $K$;
we enlarge $S$ by adding arbitrary vertices to contain exactly $K$ vertices, that is, $|S| = K$.
Construct a tree $T_1$ by interconnecting the vertices $x_v$ with $v \in S$ into a tree using $K - 1$ weight-$2$ edges and,
for every $e \in F$, connecting $y_e$ to a vertex $x_v$ such that $v \in S$ and $v$ is an endpoint of $e$, which has weight $2$.
That is, $T_1$ spans $\{x_v: v \in S\} \cup Y$ and its every edge has weight $2$, giving rise to $w(T_1) = 2(K + M - 1) = B$.
Construct another tree $T_2$ as the star centered at $v_0$ on $V \setminus V(T_1) \subset \{v_0\} \cup X \cup Z$.
All the edges in $T_2$ have weight $1$, and thus $w(T_2) = N - K + |Z| = B$ too.
This proves that $G$ a spanning $2$-forest in which both trees have weight $B$.
We prove that $\OPT = B$ in the below converse.

Conversely, let $\{T_1, T_2\}$ be a spanning $2$-forest of objective value $C = \max\{w(T_1), w(T_2)\}$ and assume that $v_0 \in V(T_2)$.
Let $d = |V(T_1) \cap (X \cup Z)|$ and $r = |V(T_2) \cap Y|$,
and $s$ be number of vertices $y_e \in V(T_1) \cap Y$ for which neither endpoint of $e$ has its corresponding vertex in $V(T_1) \cap X$.
Intuitively, $s$ is the number of edges of $F$ not covered by the vertex subset $S = \{v \in U: x_v \in V(T_1) \cap X\}$.
We distinguish two cases.

{\bf Case 1.} $d \ge 1$.
Root $T_1$ at a vertex of $X \cup Z$.
Every edge in $T_1$ has weight at least $2$, and each of the $s$ specified vertices $y_e$ has a parent edge of weight $3$ in this rooted tree.
Therefore, $w(T_1) \ge 2 (d + M - r - 1) + s$.
Root $T_2$ at $v_0$.
Every vertex of $X \cup Z$ has a parent edge of weight at least $1$ in the rooted $T_2$, and every vertex of $Y$ has a parent edge of weight at least $2$.
Therefore, $w(T_2) \ge N + |Z| - d + 2r$.
On the other hand, consider the vertex subset $S = \{v \in U: x_v \in V(T_1) \cap X\}$.
The number of edges of $F$ not covered by $S$ and their corresponding vertices in $V(T_1)$ is $s$,
and the number of edges of $F$ of which the corresponding vertices in $V(T_2)$ is at most $r$.
Therefore, $S$ can be expanded to become a vertex cover by adding one endpoint of each of the above two types of edges of $F$, leading to a size $d + s + r$.
Using $N - K + |Z| = B = 2 (K + M - 1)$, the above two inequalities on tree weights give rise to
\[
3.5 C \ge 1.5 \cdot w(T_1) + 2 \cdot w(T_2) \ge d + r + 1.5s + 3.5 B - K,
\]
and thus the vertex cover has size at most $d + s + r \le 3.5 C - 3.5 B + K$.

{\bf Case 2.} $d = 0$.
In this case, all the $N + |Z|$ vertices of $X \cup Z$ are in $T_2$.
The weight of $T_2$ satisfies $C \ge w(T_2) \ge N + |Z| + 2r = B + K + 2r \ge B + K$.

The above two cases proves the second property.
Additionally, in Case 1, the two inequalities on tree weights also give rise to
$3 C \ge 1 \cdot w(T_1) + 2 \cdot w(T_2) \ge 2r + s + 3 B$, and thus in either case we have $C - B \ge \min\{\frac 13 (2r + s), K\} \ge 0$.
This proves that $\OPT = B$ for the constructed graph $G = (V, w)$, and thus proves the first property.
\end{proof}

\begin{theorem}
\label{thm08}
For every fixed integer $k \ge 2$ and every fixed $1 \le \rho < 1688/1687$,
the {\sc BS$k$F} problem on metric complete graphs admits no polynomial-time $\rho$-approximation algorithm unless P $=$ NP.
When $k = 2$, the assertion holds even if all edge weights are in $\{1, 2, 3\}$.
\end{theorem}
\begin{proof}
Consider first $k = 2$.
Fix a rational number $\rho'$ satisfying $\rho < \rho' < 1688/1687$ and choose a fixed positive rational $\epsilon < 0.25$ such that $1 + \frac {1 - 4\epsilon}{1687 + 14\epsilon} > \rho'$.
We use a fixed number of disjoint copies of the degree-$5$ graph in \Cref{lemma22}, still denoted as $H = (U, F)$, to construct a metric complete graph $G$ as in \Cref{lemma23}.
This ensures that $K = (\mu + 4 + \epsilon) q$ is an integer.

As in \Cref{lemma23}, set $B = 2(M + K - 1) > 2 (K - 1) = 2 (\mu + 4 + \epsilon) q - 2 > 0$.
Also, $N - K = (\mu + 2 - \varepsilon) q > 0$, and $|Z| = 2M + 3K - N - 2 > 3 K - N - 2 = (\mu + 6 + 3\epsilon) q - 2 > 0$.
Thus the graph $G$ constructed in \Cref{lemma23} satisfies the two stated properties.
That is, if $H$ has a vertex cover of size at most $K = (\mu + 4 + \epsilon) q$, then $\OPT = B$.
If every vertex cover of $H$ has size at least $(\mu + 4.5 - \epsilon) q = K + (0.5 - 2 \epsilon) q$,
then any spanning $2$-forest of $G$ has objective value $C \ge \min\{B + K, B + \frac 27 (0.5 - 2\epsilon) q\} = B + \frac {1 - 4\epsilon}7 q$.
It follows from $B \le 2 (5 \mu + 17 + \epsilon) q - 2 \le (241 + 2\epsilon) q$, using $\mu \le 20.7$, that
\[
\frac {\OPT}B \ge 1 + \frac {(1 - 4\epsilon) q}{7B} \ge 1 + \frac {1 - 4\epsilon}{1687 + 14\epsilon} > \rho'.
\]
This implies a polynomial-time $\rho$-approximation algorithm for {\sc BS$2$F} would be able to solve the NP-hard vertex cover distinguish problem in \Cref{lemma22}, a contradiction.

When $k > 2$, adding $k-2$ new vertices and connecting each of them to every other vertex by an edge of weight $2B$.
This gives a metric complete graph $G'$ for {\sc BS$k$F},
that feasible solutions for $G'$ with objective value less than $2B$ one-to-one correspond to feasible solutions for $G$ with the same objective value.
Therefore, the same approximation gap holds for every fixed $k \ge 2$.
\end{proof}

\subsection{A sparse graph class}
In this subsection we show that the same lower bound of $1688/1687$ extends to a sparse graph class.

Recall that a graph is {\em $2$-degenerate} if every nonempty subgraph has a vertex of degree at most $2$,
and it is {\em minimally $2$-connected} if it is $2$-connected but the deletion of any edge destroys the $2$-connectivity.

\begin{lemma}
\label{lemma24}
Given a connected edge nonnegatively weighted graph $G = (V, E, w)$ and $2 \le k \le |V|$,
in polynomial-time a minimally $2$-connected bipartite $2$-degenerate graph $G'$ with nonnegative edge weights can be constructed such that
these two instances of the {\sc BS$k$F} problem have equal optimum values.
Moreover, any spanning $k$-forest of $G'$ can be transformed in polynomial time into one of $G$ of no greater objective value.
\end{lemma}
\begin{proof}
Denote $n = |V|$ and $m = |E|$, and $V = \{v_1, \dots, v_n\}$.
For each edge $e = uv \in E$, fix an arbitrary ordering of its endpoints, say $(u, v)$, and replace $e$ by a two-edge path $u$-$v_e$-$v$, with $w(uv_e) = w(e)$ and $w(v_ev) = 0$.
For every $i \in [n]$, add a vertex $z_i$ and two edges $v_i z_i$ and $z_i v_{i+1}$ (where $v_{n+1} = v_1$), with weights $w(E) + 1$ and $0$, respectively,
where $w(E) = \sum_{e \in E} w(e)$.
The achieved graph is $G'$, in polynomial time, and the edge-weight function is denoted as $w'$.

Since every edge of $G'$ connects a vertex of $V$ and a newly introduced vertex, $G'$ is bipartite.
The vertices $v_1, z_1, v_2, z_2, \dots, v_n, z_n$ induce a cycle, and each vertex $v_e$ is adjacent to two distinct vertices on this cycle.
This way, $G'$ does not any cut-vertex and thus is $2$-connected.
Also, every newly introduced vertex has degree $2$, and deleting an edge incident with such a vertex leaves its degree $1$, and so the resulting graph is not $2$-connected.
Therefore, $G'$ is minimally $2$-connected.
Every subgraph of $G'$ containing a newly introduced vertex has a vertex of degree at most $2$,
and a subgraph containing only the original vertices has no edges;
consequently $G'$ is $2$-degenerate.

Given a spanning $k$-forest $\{T_1, \dots, T_k\}$ of $G$,
replace each edge $e$ selected by some tree by the corresponding two-edge path through $v_e$;
for every unselected edge $e$, attach $v_e$ to its endpoint incident with the zero-weight segment;
and attach each $z_i$ to $v_{i+1}$ using its zero-weight edge.
This process gives rise to a spanning $k$-forest $\{T'_1, \dots, T'_k\}$ of $G'$, for which $T'_i$ is constructed from $T_i$, such that $w(T_i) = w'(T'_i)$ for each $i \in [k]$.
Therefore, $\OPT(G') \le \OPT(G)$.

For the reverse direction, consider a spanning $k$-forest $\{T'_1, \dots, T'_k\}$ of $G'$ of objective value $C$.
If an edge $v_i z_i$ of weight $w(E) + 1$ is selected by some tree, then $C \ge w(E) + 1$.
Computing a spanning tree of $G$ and deleting $k-1$ edges from the tree give rise to a spanning $k$-forest with objective value at most $w(E) < C$.
Otherwise no edge of weight $w(E) + 1$ is selected, meaning that each $z_i$ is either isolated or attached to $v_{i+1}$ as a leaf in some tree
(in the latter case, $z_i$ is deleted from the tree).
For each edge $e = uv \in E$, if the two-edge path $u v_e v$ is inside some tree, then the two-edge path is replaced by $e$ in the tree;
otherwise $v_e$ is either isolated or attached to one of $u$ and $v$ as a leaf in some tree (in the latter case, $v_e$ is deleted from the tree).
This process gives rise to a spanning $k$-forest $\{T_1, \dots, T_k\}$ of $G$,
for which $T_i$ is derived from $T'_i$ and specially $T_i$ denotes an empty tree if $T'_i$ is an isolated newly introduced vertex, such that $w(T_i) \le w'(T'_i)$ for each $i \in [k]$.
When there are empty trees, deleting some edges from the other trees can give exactly $k$ trees.
The objective value of the final spanning $k$-forest of $G$ is at most $C$.

This also proves $\OPT(G) \le \OPT(G')$, and thus $\OPT(G') = \OPT(G)$.
\end{proof}

\begin{corollary}
\label{coro03}
For every fixed integer $k \ge 2$ and every fixed $1 \le \rho < 1688/1687$,
the {\sc BS$k$F} problem on minimally $2$-connected bipartite $2$-degenerate edge nonnegatively weighted graphs admits no polynomial-time $\rho$-approximation algorithm unless P $=$ NP.
\end{corollary}
\begin{proof}
Using the metric complete graph constructed in the proof of \Cref{thm08} as the source graph $G$ in \Cref{lemma24},
we can construct a minimally $2$-connected bipartite $2$-degenerate edge nonnegatively weighted graph $G'$ such that $\OPT(G') = \OPT(G)$ and
every spanning $k$-forest of $G'$ can be transformed in polynomial time into one of $G$ of no greater objective value.
This proves the non-existence of polynomial-time $\rho$-approximation algorithm as otherwise contradicting to \Cref{thm08}.
\end{proof}

\section{Concluding remarks}
\label{sec:conclusion}
In this paper, we studied approximation algorithms for the {\sc BS$k$F} problem on general graphs or on metric complete graph.
On general graphs, we obtained a $(0.75 + \epsilon)k$-approximation for sufficiently large $k$, with the running time polynomial in other input but exponential in $k$ only,
and thus a polynomial time approximation algorithm for a fixed $k \ge 2$.
We also presented a $7/4$-approximation algorithm for {\sc BS$2$F}, the special case where $k = 2$.
On metric complete graphs, we achieved a constant ratio, i.e., a $3$-approximation algorithm for {\sc BS$k$F} for any $k \ge 2$, improving the previous best factor of $4$.
On the negative side, we proved the APX-hardness for metric complete graphs and for a special graph class,
by showing that in both cases, the problem cannot be approximated within $1688/1687 \approx 1.00059$ unless P $=$ NP.

A common theme behind the algorithmic results is the use of a minimum spanning tree as a structural surrogate for an optimal spanning forest.
On general graphs, contraction with respect to the optimal tree vertex sets and spanning-tree packing provide control over the weight of the minimum spanning tree,
while balanced edges and the sink auxiliary graph convert this global information into feasible forests.
In the metric setting, threshold components and a minimum-cost perfect matching play the corresponding role,
while the vertex-disjoint splitting argument is needed to preserve the partition requirement.
That is, though different, these algorithms all exploit additional structure around a minimum spanning tree to control the weight of the heaviest returned tree.

Several questions remain worth investigating.
For small fixed values of $k$, it would be interesting to improve the approximation guarantees obtained here, in particular the factor $7/4$ for $k = 2$.
For general graphs with $k$ being part of the input, a natural direction is to design an algorithm with substantially better running time dependence on $k$.
We believe it is also interesting to design improved approximation algorithms for restricted graph classes such as bounded-degree graphs, bipartite graphs,
and graphs parameterized by treewidth or vertex-cover number, etc.

\subsection*{Declaration}
The authors have no competing interests to declare that are relevant to the content of this article.

\subsection*{AI Declaration}
ChatGPT-6 (OpenAI) has been used to assist with exploring and refining some of the presented proof arguments, as well as typographical proofreading and consistency checks.
All mathematical claims, proofs, references, and the final manuscript were critically verified, re-organized and re-written by the authors.
The authors take full responsibility for the content of the manuscript.


\end{document}